# Moons Are Planets: Scientific Usefulness Versus Cultural Teleology in the Taxonomy of Planetary Science

Philip T. Metzger[1], W. M. Grundy[2], Mark V. Sykes[3], Alan Stern[4], James F. Bell III[5], Charlene E. Detelich[6], Kirby Runyon[7], Michael Summers[8].

[1]University of Central Florida. [2]Lowell Observatory. [3]Planetary Science Institute. [4]Southwest Research Institute. [5]Arizona State University. [6]University of Alaska Anchorage. [7]Johns Hopkins University Applied Physics Laboratory. [8]George Mason University.

**Abstract**

We argue that taxonomical concept development is vital for planetary science as in all branches of science, but its importance has been obscured by unique historical developments. The literature shows that the concept of *planet* developed by scientists during the Copernican Revolution was theory-laden and pragmatic for science. It included both primaries and satellites as planets due to their common intrinsic, geological characteristics. About two centuries later the non-scientific public had just adopted heliocentrism and was motivated to preserve elements of geocentrism including teleology and the assumptions of astrology. This motivated development of a folk concept of *planet* that contradicted the scientific view. The folk taxonomy was based on what an object orbits, making satellites out to be non-planets and ignoring most asteroids. Astronomers continued to keep primaries and moons classed together as planets and continued teaching that taxonomy until the 1920s. The astronomical community lost interest in planets ca. 1910 to 1955 and during that period complacently accepted the folk concept. Enough time has now elapsed so that modern astronomers forgot this history and rewrote it to claim that the folk taxonomy is the one that was created by the Copernican scientists. Starting ca. 1960 when spacecraft missions were developed to send back detailed new data, there was an explosion of publishing about planets including the satellites, leading to revival of the Copernican *planet* concept. We present evidence that taxonomical alignment with geological complexity is the most useful scientific taxonomy for planets. It is this complexity of both primary and secondary planets that is a key part of the chain of origins for life in the cosmos.

# 1. INTRODUCTION

We report the results of a five-year study of the astronomical and planetary science literature to understand how the concept of a *planet* has evolved in relationship to scientific theory and to culture. What we discovered is surprising: the story that is commonly told about the Copernican Revolution and how it changed the concept of planets is not correct.

Understanding the actual reconceptualization of *planet* is important for planetary scientists because the evolution of taxonomical concepts is a core part of science. Science creates a mental model of nature to obtain deep insight, and the model consists of both nouns and verbs: nouns are concepts like *planet*, which are organized into taxonomies, and verbs are the theories that tell what the nouns do or have done. Cognitive scientists and philosophers of science have made progress in recent decades explaining how concepts are structured in our minds (Thagard, 1990a;

Giere 1994) and how their development is operative in the scientific process (Thagard, 1990b; Craver, 2006). Healthy taxonomical concept development is central in the functioning of science, as illustrated in Fig. 1. Our minds organize empirical data into taxonomical concepts that are aligned with theory, and those taxonomies organize our efforts to abductively create and evaluate new hypotheses leading to better theory. Thus, the taxonomical concepts and the theories co-evolve.

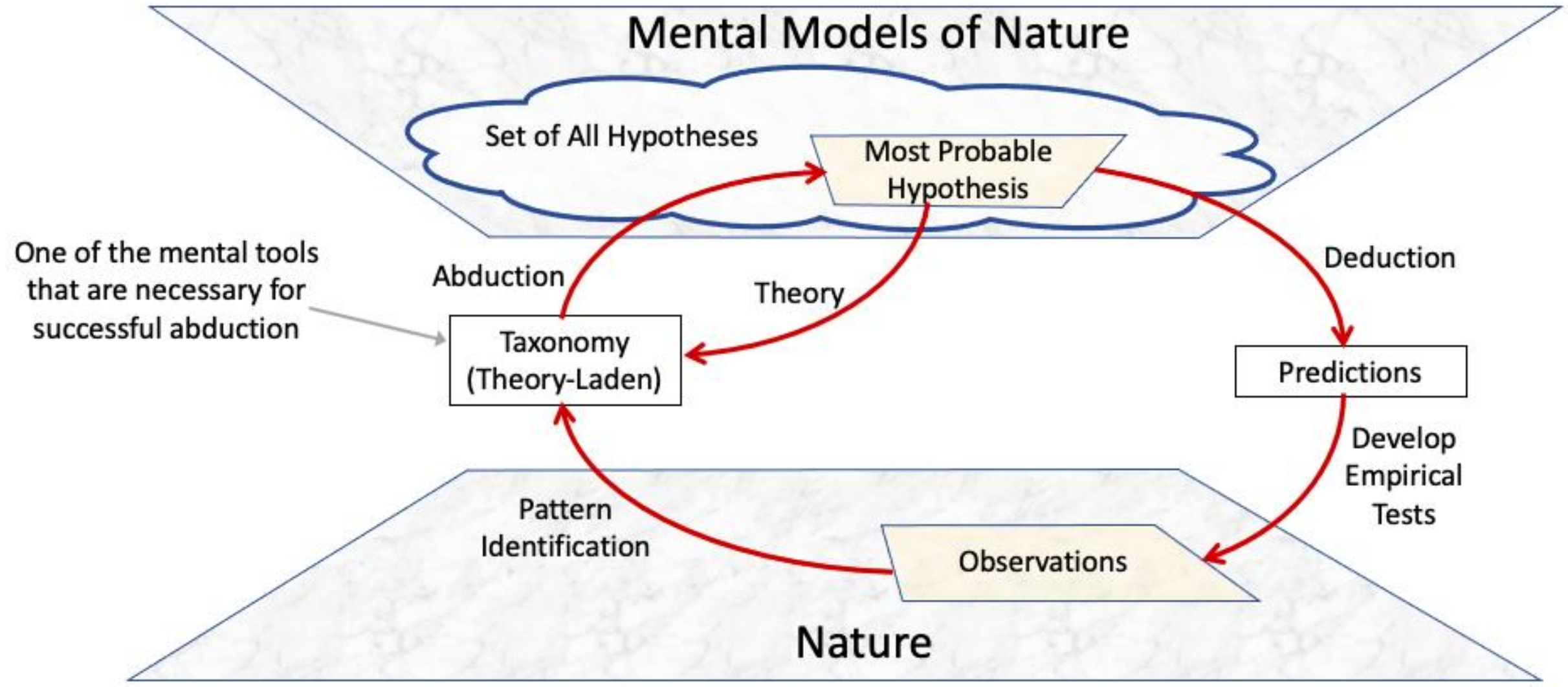

**Figure 1.** Flow diagram of a model of the scientific method showing taxonomy's functional role. Portions modified from Fossion and Zapata-Fonseca (2018), Fig. 2.1.

Cognitive scientists argue that everything our mind does is theory-laden (e.g., Estany, 2001). Taxonomies are aligned with theory and reflect the structure and organization of the theory, even while the theory is provisional; there is no such thing as a naked observation of nature without many layers of theory (Gelman and Coley, 1991; Pearson, 2010; Fitzhugh, 2013; Wilkins and Ebach, 2013; Hazen, 2019). "[E]xplanatory, causal, teleological, and ontological beliefs and biases…condition almost all category-based inferences" (Rehder and Hastie, 2001). Taxonomy is theory-laden not only in science but in everyday cognition, too, although the theories that scientists build into taxonomies generally have different goals than the theories important in everyday cognition. For example, when scientists insisted that dolphins are mammals, not fish, it communicated insight into the organization and history of the natural world by directing us to the theory that aligns with the taxonomy. A *planet* in science is therefore not an arbitrary concept but is a class of objects defined by the community of practice as they repeatedly perform abduction to develop improved explanatory theories of those types of objects, grouping or ungrouping the objects to make the class align with the co-evolving theory.

Taxonomy is both an important outcome of our reasoning processes and a functional cog that enables those processes. This is true in scientific cognition as well as in everyday cognition. Taxonomy motivates and aids the abduction of new hypotheses and the evaluation of hypotheses. Thus, "scientific knowledge can be understood as both subject and object of the continuous revision of categories" (Freeman and Frisina, 2010). Philosophers of science have described our ability to abductively create hypotheses as seemingly a "miracle" (Pierce, 1935) because we can leap to new causal connections where straightforward reasoning is impossible. Research into abduction is an active field making rapid progress, and we think the findings underscore the importance of taxonomy. Artificial intelligence researchers find that it is easy to write algorithms to do logical deductive reasoning, but it is beyond our current capability (apart from limited cases) to write algorithms that do successful abductive reasoning (Bylander et al., 1991; Psillos, 2000; Creignou and Zanuttini, 2006; Nordh and Zanuttini, 2006; Hermann and Pichler, 2010). Many tricks are employed to make abduction tractable (e.g., Josephson et al., 1987; Eiter and Gottlob, 1995; Gottlob et al., 2007; Feldman, 2010), and these are presumably analogous to the methods the human brain has adapted. For example, setting limits on the size of composite hypotheses is necessary to make abduction computationally tractable (Bylander et al., 1991). Johnson-Laird (2010) wrote that "almost all sorts of reasoning…are computationally intractable. As the number of distinct elementary propositions in inferences increases, reasoning soon demands a processing capacity exceeding any finite computational device, no matter how large, including the human brain." By creating taxa that align naturally with theory, we cut out unnecessary concepts and avoid composite hypotheses from amalgamations of poorly aligned concepts. Hence, the development of taxonomical concepts that align naturally and deeply with theory appears to be crucial for effective reasoning in all situations, not just specific scientific fields like zoology. As we argue here, planetary scientists have defaulted to doing useful taxonomy in recent decades despite the strong presence of a non-scientific folk taxonomy, and despite the International Astronomical Union (IAU) mistakenly embracing that folk taxonomy. We think that this strong pull toward functional taxonomy despite these hinderances demonstrates that taxonomy is central and necessary in human reasoning.

When we look at the way a community has defined its taxa, we learn something about the values held by that group because the taxa align with the theories they were co-evolving with them and therefore they reveal what problems the community believed were important to solve. The public defines fruit as an opposite to vegetables because they value the eating experience and need taxonomical concepts to discuss the selection of food, but plant biologists define fruit as reproductive organs in plants because they value explanations of nature. Therefore, a green bean is a fruit to biologists but not a fruit to the public. Both taxonomies are equally valid for their own purposes, both are theory-laden as mental constructs according to cognitive scientists, but both are not equally useful to explain the evolutionary and ecological behaviors of plants. As we argue below, the folk taxonomy of our Solar System, which sees planets as an orderly system of just a few culturally significant objects, reflected the values and cosmological theories of the non-scientific public during the early 1800s when it developed.

Polysemy is the property of words to have different but significantly overlapping meanings (Nerlich, et al., 2011). Biologists have argued over polysemic meanings and definitions for *species* since at least the 1700s (Hey, 2001; Wilkins, 2018). Brusse (2016) argued that the polysemy of the word *planet* existed undetected in the astronomical and planetary science communities for a long time, and its sudden exposure in 2006 resulted in a dispute of surprising intensity and duration. Messeri (2010) argued that the polysemy of *planet* had functioned as a beneficial "trading zone" that enabled different subcommunities of astronomers and planetary scientists to discuss research without need of a precise definition, but that the IAU's 2006 *planet* definition disrupted the trading zone. Polysemic meanings for *planet* vary not just between subcommunities of scientists but also between scientists and the public. Owen Gingerich, who was the chair of the IAU's Planet Definition Committee, wrote in 2007 that it was a mistake to have attempted defining *planet* because it is "too culturally bound, with elastic definitions that have evolved throughout the ages" (Gingerich, 2007). We agree it was a mistake at that time, so studies such as this one are needed to untangle the issues, enabling scientific progress. However, in 2008, Brown (2008) argued that scientists can lay out the options for a *planet* concept, but there are no scientific reasons why scientists might choose one over another, so the choice is "purely cultural"; to argue that there are scientific reasons for a planet definition is "conflating the job of science and the job of culture." Many other astronomers demonstrated in 2006 their belief that defining *planet* is essentially cultural by arguing that the IAU should create a definition that matches (what they thought was) the public's expectation of a small number of planets so the public would accept it. We did not find a record of any discussion leading up to the IAU's vote regarding *why* the public thought planets should be few in number, whether the theory that lay behind such a taxonomical concept was something the public *ought* to hold or ought to reject, nor the reasons why scientific and folk taxonomies must oftentimes disagree. There is a deep literature on the interaction of scientific concepts and competing folk concepts (also called *intuitive ontologies*). For example, folk taxonomies for animal and plant species are globally ubiquitous (Medin and Atran, 1999), and they are studied to understand how they developed and how they are structurally like and unlike the corresponding scientific taxonomies (e.g., Atran, 1998). As we argue below, this idea that *planet* is different than the concepts of other branches of science such that its definition is not "science" and therefore should be determined essentially by the culture is misguided and is the result of the historical presentism fallacy that has confused modern astronomers.

Scientists have fought against folk taxonomies in other branches of science, for example the case with whales and fish (Petto and Mead, 2009; Burnett, 2010). The public has finally agreed, largely abandoning the old folk taxonomy in which everything living in the water was a fish (jellyfish, starfish, shellfish, cuttlefish, whales, etc.) and broadly accepting into everyday usage a taxonomy aligned with evolutionary relationships. This represented an infusion of scientific understanding into culture. Other folk taxonomies are tolerated in co-existence with the scientific ones because they do no harm scientifically and they do have cultural value: the two meanings of "fruit" are an example. Although they co-exist, scientists do not allow the folk *fruit* concept to replace the scientific one for publishing about plant reproduction, and when scientists explain to the public how they have a different definition of *fruit* it becomes an opportunity to teach biology. On the other hand, De Cruz and De Smet (2007) warned that "scientists exhibit the same cognitive biases and limitations as other human beings…If intuitive ontologies continue to guide our everyday understanding, there remains the possibility that intuitive ontological ideas may slip unnoticed into scientific discourse."

The story of folk versus scientific taxonomies for *planet* that we present here is far more complicated than the *fish* or *fruit* examples, and it is a much larger issue than many realize; it is about the need to clearly demark scientific pragmatics from cultural priorities so that the deepest scientific insights will be made crystal clear. Restoring that demarcation in planetary science will require an understanding of how we got to this point, why it is important to repair, and what is a better path forward.

The first part of this paper (sections 2 through 5) will attempt to unravel this history to identify the competing theories that drove the development of the folk and scientific taxonomies for planets. The second part (sections 6 through 8) will describe how planetary scientists are using the *planet* concept in pragmatic, scientific usage today, despite the IAU's voted definition. This will help us identify the modern theory that is driving the pragmatic taxonomy. It will demonstrate the deeply insightful nature of the scientific *planet* concept in contrast to the less scientifically insightful concept chosen by the IAU.

Section 2 sets the stage by briefly recounting the events surrounding the IAU's vote in 2006. Section 3 then backs up in history to demonstrate what happened in the taxonomy of planets among scientists beginning with the Copernican Revolution until the 1920s. Section 4 shows what happened to the *planet* concept among the non-scientific public as they resisted Copernican cosmology then in the 1800s sought accommodation within heliocentrism for a non-reductionist perspective carried over from geocentrism. Section 5 analyzes the bibliometrics and literature of the period 1910 to 1955 to explain how and why astronomers abandoned the Copernican taxonomy and adopted the folk taxonomy at that time. Section 6 shows that the modern usage by planetary scientists calling moons *planet* since about 1960 is a rediscovery of the essential insight of the Copernican Revolution but now with far greater theory and data to support it. Section 7 presents additional data demonstrating that planetary science is broadly based on a geological/geophysical concept of *planet* contrary to the IAU's definition that was based on the folk taxonomy. Section 8 argues that the geological/geophysical conceptualization of planets, both Galilean and modern, is one of the most important concepts in all scientific history because it aligns with an important reductionist explanation in the "great chain of origins", helping us to understand our place in the cosmos. In Section 9 we conclude that it is important for planetary scientists to reject the folk taxonomy and embrace the geological/geophysical one.

# *2.* The IAU's 2006 Vote on a Definition of *Planet*

By the late 1990s, new technology enabled astronomers to discover many new Kuiper Belt Objects (KBOs). The Minor Planet Center, commissioned by the IAU, named the new KBOs routinely using the same procedure that it uses to name asteroids and comets. When 2003 UB313 (later named Eris), a body close to the size of Pluto, was discovered, the question arose whether it was a "planet" and whether it could be named by the existing procedure for small bodies. No large, primary planet had been named since Pluto, which occurred in 1930 when asteroids were still considered to be planets (Metzger et al., 2019), so separate naming processes for "planets" versus small bodies had never been established. The IAU leadership thought that to answer the naming question they first needed to decide whether large KBOs are "planets" or not, essentially picking sides in a taxonomical matter to resolve a bureaucratic matter.[1] This created a sense of urgency in settling a scientific taxonomy. Ron Ekers (2018), who was IAU president at the time of these events, wrote that "no name could be assigned by the IAU because the naming conventions are different for minor solar system bodies and for planets." He later elaborated:

[1] Ekers (2019) argued during a debate that defining "planet" was not science but was just naming. We disagree for the reasons given in Section 1 and because the IAU's resolutions in 2006 established the ranks and memberships of

> Now IAU agrees on the names that are given to new objects that are found in the sky. However, there was no definition of whether it [2003 UB313] was a planet or not, and that meant that the process for naming it could not be continued because you had to first make a decision of whether it was going to be called a planet, in which case it is named after a major god, or not a planet and then it is not [named after a major god]. All of this seems ridiculous, but this is the part of the conventions agreed on for naming things.[2] (Ekers, 2019)

We believe this was a grave error because the processes simply should have been changed to enable naming objects without settling their taxonomy; taxonomy should never be settled by voting for the reasons discussed in Section 1. We think most astronomers were unaware of the need to separate taxonomy from naming in part because their views on taxonomy had been skewed by the strong historical presentism fallacy. Historians define presentism as the erroneous attribution of modern perspectives to people who lived during an earlier era when those perspectives had not yet come into existence (Fischer, 1970; Hunt, 2002). We argue in this paper that astronomers have erroneously attributed an 1800s folk taxonomy, one that is not aligned with reductionist or ontogenetic theory and therefore has low value for science, to the Copernican Revolution. That false pedigree made it seem normal and legitimate to have a non-functional taxonomy for planets, and that led many astronomers to think that comingling taxonomy with naming is acceptable. The rushed effort to settle the naming procedure for 2003 UB313 prevented these issues from coming to light in time to affect the 2006 discussion.

We think there was another grave error when the IAU leadership decided to proceed with a vote in 2006: there was inadequate consensus on the definition of a planet. Just prior to the vote, the IAU prefaced it by writing, "…IAU recommendations should rest on well-established scientific facts and have a broad consensus in the community concerned" (IAU, 2008). The IAU's Bye-Laws and Working Rules were written to protect that consensus. The Bye-Laws say,

> (7) d. Any motion of a scientific character…shall be placed on the Agenda of the General Assembly, provided it is submitted to the General Secretary, in specific terms, at least six months in advance.

---

several taxa, which is not naming. The zoological community, for comparison, uses a naming process that avoids settling the ranks and memberships of taxa, a process that they developed for the explicit intent of preserving the taxonomical freedom of scientists (Ride, et al., 2012).

[2] This greatly overstates the problem. The counterproposals in 2006 would have made either a non-god (Charon) into a major planet or a major god (Pluto) into a non-planet, so it was never an option to align the taxonomy with mythological significance. The chaotic histories of naming Uranus and Neptune demonstrate there was no naming convention during those periods. Several of the asteroids were named after gods that are equally or more significant than Mercury (e.g., Ceres and Juno); perception about Mercury being a major god seems biased by its status as a major planet. During that period when asteroids were accepted as full-fledged primary planets, some asteroids (e.g. 18 Melpomene) were named for mythological beings that are not gods at all. The dwarf planets Haumea and Makemake were named by the IAU in 2008 after they had been declared non-planets, but those are the names of two of the most major gods of their respective cultures. Many other asteroids and KBOs today are named after major gods from non-Greek and non-Roman mythologies. We think the sense of urgency in 2006 did not really come from the question of major gods, which has been so easily set aside and necessarily so, but rather from the procedural question whether the small bodies committee had authority to decide the names of planets versus the total absence of another procedure if it did not.

> e. The complete agenda, including all such motions or proposals, shall be prepared by the Executive Committee and submitted to the National Members at least four months in advance.[3]

The IAU's Working Rules are written by the executive committee without approval of the membership so they can apply but not override the Bye-Laws. They say,

> Resolutions should be adopted by the Union only after thorough preparation by the relevant bodies of the Union. The proposed resolution text should be essentially complete before the beginning of the General Assembly, to allow Individual, Junior and National Members time to study them before discussion and debate by the General Assembly. The following procedures have been designed to accomplish this:
>
> (…) Upon submission each proposed Resolution is posted on the Union web site…. Resolutions must be submitted to the General Secretary six months (Bye-Laws 7c and 7d) before the beginning of the General Assembly. The Executive Committee may decide to accept late proposals in exceptional circumstances.

The clause allowing the Executive Committee to accept late proposals is not permitted by the governing regulations (the Statutes and Bye-Laws) so it represents a "bending of the rules" that the Executive Committee thought was reasonable within the context of the other things they said about thorough preparation and months of review ahead of the General Assembly (GA).

In 2004, the IAU began the process to develop a definition when it established a Division III (Planetary Systems Division) Working Group. This group was unable to find consensus (Boss, 2009; Schilling, 2009). Ekers stated that this failure was not due to the scientific questions but "aspects that were related to social and cultural issues…" (Ekers, 2006). In early 2006 the IAU executive leadership sought to address those broader issues by establishing the Planet Definition Committee, which now also included historians, writers, science communicators, and educators. The IAU executive leadership also decided to set aside the Bye-Laws and the Working Rules when they directed the new committee to keep its discussions secret and not to reveal the proposal until after the GA began. As a result, the IAU did not circulate the resulting proposal for the required four months nor did they publish it on their website. Ekers (2018) later said,

> Normally, resolutions to be voted on are formulated and made public many months ahead of a GA but, given our awareness of the sensitivity and the risk of derailing the process, we made the decision not to announce anything about the planet definition resolution before the first business meeting of the GA in Prague. Although this ran counter to the need to strengthen the involvement of the

[3] The Bye-Laws had been changed in the 2003 General Assembly (GA) to eliminate individual voting on scientific matters, so all votes would be cast by the national members (i.e., national organizations) as proxy for their own individual members. This was immediately considered a mistake, so it was rescinded during the 2006 GA the day before the planet definition proposal was revealed so that individual members could vote on the planet definition. However, the rules both before and after these changes required that proposals on scientific matters be made public months in advance to enable adequate review by scientists to protect consensus.

> membership and the need to conduct all IAU business in an open and transparent manner, we knew that once we made it public there would be strong media interest and pressure to influence the decision-making process.

He later added,

> Now, here is one decision where I think you can very much question what the IAU did. We decided not to discuss this in public before the meeting in Prague…That, I am willing to accept, may have been a wrong decision. (Ekers, 2019)

The Planet Definition Committee was apparently unaware that the IAU leadership was violating its Bye-Laws, and they developed a good-faith proposal that we believe was a step in the right technical direction to restore functional taxonomy to planetary science (although we now think that voting on it was inherently a mistake). The proposal was met with anger mainly by the dynamicist subcommunity because the proposal favored a geophysical *planet* concept over either a dynamical one or a "balanced approach" that included both aspects of planets. Ekers (2019) reported,

> [O]ne remark I remember quite clearly [from the dynamicists] was, "If you don't include the clearing of the orbit in the definition, you have insulted half of the planetary science community."

The use of loaded terminology ("insulted") and the emotional reactions described by witnesses (Schilling, 2006; Boyle, 2009) are evidence of the social dynamics that the scientific process normally tries to avoid because they interfere with reasoning (Oaksford et al., 1996; Dolcos and McCarthy, 2006). The angry responses are also a demonstration that there was no consensus.

We think one of the reasons there was no consensus was because the issues that we now discuss in this paper have been largely unknown to the modern astronomical community. Driving out these types of issues takes a lot of time – decades or longer. For comparison, biologists have been arguing over a definition for *species* since the 1700s or earlier, which has helped clarify the issues and supported progress in biology. We think the IAU leadership's decision to force a vote on the *planet* concept with only a few days of debate exacerbated deep divisions by forcing scientists to pick sides prematurely and to participate in an emotional social dynamic that created factions and a psychological need to defend individual and organizational pride. Now consensus may be even more difficult to obtain because this has likely activated and amplified a variety of cognitive biases (Bénabou and Tirole, 2016) including post-purchase rationalization (Lind et al., 2017), groupthink (Janis, 1971), self-serving bias (Forsythe, 2008), false consensus effect (Ross et al., 1977), status quo bias (Kahneman et al., 1991), asymmetric updating of beliefs (Sharot and Garret, 2016), system justification (Jost et al., 2004), and conformity bias (Baron et al., 1996).

In a period of about four days during the GA, a group led primarily by dynamicists began meeting and wrote an alternative definition then pushed it through to a vote. Apparently, they believed that this action was allowable because it seemed no different than what the IAU leadership had just done to them, surprising them with a proposal on the second day of the GA with no prior chance to analyze and assess it. Also, the executive committee, which should have known the extent to which the rules were being broken, did not step in to stop it, furthering belief in the legitimacy of those actions.[4] Only a little more than 400 members were still present to vote at the end of the GA (Schilling, 2010). Planetary scientists, especially planetary geoscientists, were deeply non-represented partly because they are not members of either the IAU or its member organizations; partly because planetary scientists do not ordinarily attend the IAU GAs as it is not a venue where much planetary geoscience takes place; partly because the existence of a planet definition proposal was kept secret so the community was not alerted to participate; and partly because active resolution-crafting is not supposed to take place during the GA so they could not have known how important it would be to participate. The arguments from both sides during the GA were heavily invested in how the public would respond to the definition since planets are important in general culture (Anon., 2006; Messeri, 2010) but we found no records or reports of any discussions addressing the issues that we present in this paper. Some argued that the public would not accept a definition that allowed more than a small number of planets. Others were concerned with creating a definition that represents the dynamicists and the geoscientists equally so as not to "insult" any part of the community.[5] When the vote was taken it again demonstrated a lack of the broad consensus to which the IAU aspires, since the votes were split approximately 2/3 for and 1/3 against, according to some witnesses, or a narrow majority according to others.[6] This showed that it was not at a level that could be considered consensus. The strength of the vote against the resolution is surprising considering the relative absence of the community most likely to be against it.

The resulting definition embraced by the IAU has three parts:

> A planet is a celestial body that
> (a) is in orbit around the Sun,

[4] Statements in the 2006 GA newspaper show that the Executive Committee expected there would be at least some revision to the proposal during the GA, which would be consistent with both the letter and the intent of the Bye-Laws. Ekers (2019) and others have described how they did not expect such a strong backlash from the proposal and how they realize now that they were mistaken in the way they brought it to the GA. Those comments reveal that they did not expect the GA would go so far as a wholesale creation of a new proposal. In other words, they did not expect that their measured violation of the Bye-Laws, keeping the months of work a secret until the GA, would escalate into a further violation, writing a completely new resolution on a controversial matter during the course of the GA, itself. In any case, we know that the leadership embraced the ensuing process because the proposal committee reviewed the new proposal during the GA and gave approval to take it forward to a vote. We think that this setting aside of the Bye-Laws, both by keeping the original proposal secret and by allowing a new proposal to be crafted on-the-fly, reflects the leadership's belief that the most important issue was not following the rules designed to protect consensus (on a matter that they did not consider to be real science [Ekers, 2019]), but deciding how to name the KBOs, which they believed was urgent.

[5] We note that this combination of different types of properties, both dynamical and intrinsic properties, in the same rank of a taxonomy is a violation of basic taxonomical principles (Lazarsfeld and Barton, 1955), like mixing a definition of *carnivore* with a definition of *mammal* in the same rank of a single taxonomy. These different aspects should be the subjects of distinct taxonomies or of different ranks in a taxonomy. We found no record of any discussion of these types of taxonomical issues at the GA. This is another symptom of the lack of adequate time to deal with a complex matter.

[6] Ekers (2018) reported that it "passed comfortably with no need for a count," so they made the decision in real time not to count the vote. Gingerich (2014), who watched the vote by livestream, reported it was "on a knife edge balance, like approximately six votes in 400 or so votes being cast." Analysis of the photographs of the vote suggests it was not an overwhelming majority.

(b) has sufficient mass for its self-gravity to overcome rigid body forces so that it assumes a hydrostatic equilibrium (nearly round) shape, and
(c) has cleared the neighbourhood around its orbit.

(IAU, 2006)

While this purports to be an *intensional* classification by listing requirements for membership, its interpretation is restricted by a footnote that is an *extensional* classification by listing all the members (Marradi, 1990): "The eight planets are: Mercury, Venus, Earth, Mars, Jupiter, Saturn, Uranus, and Neptune".[7]

As we argue in this paper, the idea that the planets must be few, monocentric, and orderly was not developed by scientists to cohere with any theory. It developed in the 1800s as the public was converting from geocentrism to heliocentrism and they smuggled ideas from geocentrism into their new worldview. We believe that the scientists in 2006 who supported the IAU's definition on the grounds that the culture would not accept too many planets were badly underestimating the willingness of modern culture to embrace good, scientific taxonomy even if it means having many more planets. The culture of the early 2000s is no longer the culture of the early 1800s, and the conditions that created the expectation of a small number of orderly planets no longer exist. We find that the members of the non-scientific public who support the IAU's definition do so because they were told by the IAU that it represents good science, and on the IAU's authority they want to embrace good science. We find there are many others who reject the IAU's definition because they believe (as we do) that it is bad science. Both groups want good science. We think the IAU's desire in 2006 to keep the number of planets small was therefore not only bad science but also a poor appraisal of modern culture.

Some astronomers have subsequently claimed that the IAU's definition is useful for shaping the views of the public about planets because (they claim) the idea that some of the Solar System bodies are able to achieve a metastable dynamical arrangement for a relatively long time is the most important thing the public needs to understand about planets (AAAS, 2014). The IAU definition has now been adopted into science textbooks, which claim that astronomers created the definition based upon advances in science. This has been rejected emphatically by many planetary scientists for four reasons: first, because it does not represent the way the planetary science community uses the *planet* concept in actual science (see Sections 6-8); second, because it was rushed (Ekers, 2018) and therefore it was both imprecise (Soter, 2008) and narrowly focused on only one solar system (Grinspoon, 2013) and therefore not useful in science even if it had been aligned with scientific usage; third, because taxonomical concepts like *planet* should never be voted upon because taxonomy is a vital part of the scientific process (Ghiselin, 1969; Freeman and Frisina, 2010; Fitzhugh, 2013; Hazen, 2019), so having authoritative bodies decide on taxonomy subverts science (Ride, et al., 2012); and fourth, because it represents to the public that taxonomy can be decided by voting, which undermines their understanding of the scientific process and may undermine their trust in the scientific endeavor, producing long-term harm to science (Sykes, 2008; Stern, 2011; Metzger et al., 2019). We further make the claim here that the view of planets as engines of complexity up to and including life, regardless of their current dynamical state, is the far more important thing that the public needs to understand about planets, and this is consistent with both historical and modern scientific usage. The dynamical orderliness of a dominant subset of these bodies is also important but far less so. The public has always erred on the side of presuming orderliness in the heritage of geocentrism, so it needs to learn the balancing truth about the chaotic and contingent nature of stellar system architectures that extend beyond our limited human perspective.

[7] We note that intensional and extensional classifications are incompatible because they serve opposite purposes in scientific operation (Marradi, 1990). Because the final definition included both, neither purpose can be operational. We think this is a symptom of the IAU's mixture of taxonomy with naming. Taxonomy is inherently non-arbitrary and open-ended to support scientific progress, so we expect intensional classifications should be used, reflecting the theory-ladenness and scientific pedigree of the evolving concept. Naming is inherently arbitrary (not theory-laden) and requires closure. We think that the decision to include both types of classifications, with the arbitrative piece lowered into a footnote making it appear as not the main piece, happened because the group crafting it wanted the definition to present itself to the public as having the scientific stature of an intensional classification yet give the IAU's naming process the complete closure of an extensional classification.

Some astronomers have claimed that the third part of the IAU's intensional definition has historic precedence because asteroids were once planets until they were (supposedly) removed from planet-status in the mid-1800s after astronomers realized there are 15 or more of them sharing similar orbits. As mentioned above, this is not based upon fact but is a presentist revision of history, unrelated to the actual scientific treatment of asteroids (Metzger et al., 2019). In this paper we now deal with the first part of the IAU's definition, the idea that planets only include bodies that directly orbit the Sun (or by extension another star) so satellites are not planets. We address the claim that some astronomers are making that this was historically established since the Copernican Revolution. We show that to the contrary, explanatory science has always included satellites as planets for vital reasons.

# 3. SCIENTISTS' CONCEPT OF PLANET, 1610 – 1920s

In Europe prior to the Copernican Revolution, the Moon and the Sun were two of the seven known planets (Hall, 1971; Heather, 1943; Lewis, 2013; Michaelis, 1982). After the Copernican Revolution, scientists no longer considered the Sun to be a planet because it became one of the "fixed stars", and the Earth became a planet because it moves, but to the Copernican scientists the Moon *remained* a planet and was joined by other moons as planets. Having status as one of the main bodies in a primary orbit was not a part of the *planet* concept among scientists from the beginnings of heliocentrism until the 1920s. The literature shows unmistakably and consistently that this was the broad consensus of astronomers in the Latin west. We studied Latin, English, French, German, and Spanish-language publications. We found that some of the terms in other languages were different, as expected (*Neben planeten* in German vs. Secondary Planet in English; *planetoid* in German vs. asteroid in English), but they were applied to the same concepts with the same usage across all the literature we surveyed. Due to space limitations, we provide only a few quotations as examples from the literature, but more extensive documentation is provided in the supplementary material.

## <u>3.1 Early Theoretical Coherences with the Taxonomy</u>

The concept of planets that included both primaries and moons was not arbitrary but was a theory-laden taxonomy that scientists needed. At the Copernican Revolution, if scientists had been developing theories that moons and primaries possessed different properties correlated to or resulting from their different orbital states, they might have created a taxonomy in which the orbital state was the most important feature of Solar System bodies and hence moons would have been defined as non-planets. Instead, the literature shows that scientists were interested in theories that depended crucially on moons and primaries sharing common properties, so the orbital hierarchy of the Solar System was put into a lower tier of the taxonomy.

Galileo observed geological features on the lunar surface and argued on their basis that the Moon is a geological and changing body like the Earth, not made of unchanging ether as in Aristotelian philosophy (Galileo, 1638; transl. by Fabbri, 2012; Cohen, 1993; Shea, 2000).[8] From his comparison of the Earth and the Moon he extrapolated that all the planets must be changing, geological bodies, or "other Earths". This indicated that there is no perfect, unchanging celestial physics governing the planets distinct from the earthly physics governing the Earth; earthly physics is in the heavens, too, and hence the Earth is in the heavens and moving like the rest. This argument depended on the observed geology of two planets, one being a satellite (the Moon) and the other being a primary (the Earth). It would have undermined the structure of this argument, and thus the promotion of heliocentrism, to say moons are not a member of the class of bodies Galileo was arguing about. As Galileo noted,

> …many detailed parallels were drawn between the earth[9] and the moon. More comparisons were made with the moon than with other planets, perhaps from our having more and better sensible evidence about the former by reason of its lesser distance. (Galileo, 1632)

Galileo further argued that not only do the planets have the geological properties of Earth, but the Earth has the luminous property of the planets. Geocentric astronomy held that stars were either wandering stars (the seven planets) or fixed stars (what we now simply call stars), meaning they were in fixed relative position to each other. In geocentrism the planets were diverse in their intrinsic properties: at least one was opaque but reflective (the Moon) and at least one was self-luminous (the Sun), while the source of light from the other five planets was unknown (Fabbri, 2016). Galileo argued for the opacity of the Moon (some contemporaries held it was translucent), and he observed the phases of Venus, proving the latter is also an opaque but reflective body. He further observed the reflection of Earthshine off the Moon, which he argued is proof that the Earth would appear as a brilliant "star" just like the Moon and the other planets if it could be seen from a distance, even though the Earth is obviously an opaque and not self-luminous body beneath our feet. From the shared property of reflective opacity of these three bodies (the Earth, the Moon, and Venus), Galileo argued reductionistically that in heliocentrism the stars' intrinsic and dynamical classifications collapse into a single classification: all the wandering stars (the planets) are the reflective and opaque ones, while all the fixed stars are the self-luminous ones (Galileo 1610, 1611):

---

[8] Our focus is to describe the main themes of taxonomical development beginning with the Copernican Revolution to show the taxonomy was theory-laden. We do not attempt writing a complete history, nor do we survey taxonomical developments before the Copernican Revolution. See, e.g., Gingerich (2011), Fabbri (2012), and the references therein for historical analysis of Galileo's arguments including their context, relationship to Plutarch, etc.

[9] As proper names, Moon, Earth and Sun should be capitalized, and we do so consistently in our own writing. When quoting documents that did not capitalize them, we keep the quotations exactly as found in their original or as provided by the cited translation. Older English documents often used variant spellings and capitalized many nouns that are not proper nouns, so we kept those as per the originals.

> I have most conclusive arguments ready, showing clearly that… all the planets receive their light from the sun, being by constitution bodies dark and devoid of light; but that the fixed stars shine by their own proper light… (Galileo, 1611)

Because the Earth is an opaque but reflectively brilliant star like the planets, Galileo argued that the Earth must share the other properties of planets, namely their motion relative to the fixed stars, and so the Earth moves. Again, the form of his argument depended critically on the most observable planet, the Moon, sharing class similarity with the other planets. This argument like the first one shows that the taxonomy was not arbitrary but was theory-laden and pragmatic to promote science. Galileo's geophysical insight about opacity versus luminosity may seem pedestrian to the modern scientist, but it was remembered and repeated as the defining property of planets for centuries, e.g., Keill (1730), p. 91; Ferguson (1753), p. 9; Morse (1814), p. 15; Wilkins (1841), p. 4; Mc'Intire (1867), p. 28; Leahy (1910), p. 84.

When Galileo found the moons of Jupiter, he and other astronomers classified them as planets:

> Four Planets revolving around the star of Jupiter…I should disclose and publish to the world the occasion of discovering and observing four PLANETS, never seen from the very beginning of the world up to our own times… These are my observations upon the four Medicean planets… (Galilei, 1610, all caps in the original)
>
> …for now we have **not one planet only revolving about another**,[10] while both traverse a vast orb [orbit] about the Sun, but our sense of sight presents to us four stars[11] circling about Jupiter, like the Moon about the Earth…(Galileo, 1610, bold added)

In contrast to this, geocentrism held that the Earth was the lowest location in the cosmos and therefore the unique center of planetary orbits. It was an inherently monocentric system. The existence of planets centered on Jupiter supported heliocentrism by proving that planets do not need a cosmologically unique center for their orbits (Robinson, 1974). If Copernican astronomers had classified the satellites of planets as non-planets, it would have weakened the polycentric planets argument that they were making and thus undermined their case for the theory of Copernicanism. Huygens (1659) and Cassini (1673) classified the moons of Saturn as planets. Herschel (1787) classified the moons of Uranus as planets. The literature shows there was universal agreement among heliocentrists that moons are planets and that this taxonomy co-evolved with the theory of heliocentrism for pragmatic benefit (e.g., Kepler, 1620; Wing, 1654; Brigden, 1659; Huygens, 1659; Cassini, 1673; de Fontenelle, 1686; Huygens, 1698; Newton, 1692/3; Cotes, 1712/3; Gregory and Halley, 1715). See further examples in the supplementary material. The planethood of the Moon and other satellites was not a throwaway oddity of the period; it was central to the scientific reasoning of the period, and the Copernican Revolution cannot be properly understood without this conceptualization of planets.

---

[10] We added bold to some quotations as a guide to the eye to identify the portion that demonstrates the point. Where we have done so we have noted it in parentheses following the quote.

[11] Phenomenologically, wandering stars (planets) and fixed stars were all still called *stars*.

## 3.2 Dynamical State is a Subclassification

While the main themes of planetary theory benefitted by keeping moons as planets, there are of course reasons why scientists must refer a planet's current dynamical state, so Kepler invented the terminology *primary planet* and *secondary planet* to distinguish primaries from moons. This division was a lower rank in the taxonomy than the *planet* taxon so they defined subclasses of planets (Whiston, 1717; Treiber, 1719; Pemberton, 1728; d'Alembert, ca. 1765; Strong, 1784; Cousin, 1787; Herschel, 1787; Du Séjour, 1789; Hershel, 1795; Woodward, 1801; von Freygang, 1804; Funk, 1837; Dick, 1838; de Pontécoulant, 1840; Wilkins, 1841; Cornwell, 1847; Crampton, 1863; Gavarrete, 1868; Kiddle, 1870; Vaughan, 1871; Nasmyth and Carpenter, 1874; Lynn, 1891; Gore, 1893; Doolittle, 1901; Todd, 1901; Chambers, 1911). See supplementary material for more examples. Leahy wrote the following example in 1910, showing that the *planet* concept was still unified by geology/geophysics even across differing dynamical states:

> The **satellites or secondary planets**, over twenty in number, are all too feeble gravitationally to retain at their surface a proper atmosphere…**True planets all of these are** whatever be their distinguishing names [i.e., primary vs. secondary]. For they are solidified bodies shining by reflected light. (Leahy, 1910, bold added)

Furthermore, with few exceptions the early scientists did not say that primary planets are the only ones that orbit the Sun: they were careful to say that moons do orbit the Sun, too, but they do so as companions of a primary giving them a compound orbital motion (Galileo, 1610; Kepler 1620; Dick, 1838; Wilkins, 1841; Mattison, 1851; Smith, 1856; Gavarrete, 1868; Kiddle, 1870). Thus, secondary planets follow both primary and secondary orbits, whereas primary planets follow only primary orbits:

> THE true Solar system, or, as it is sometimes called, the Copernican system, consists of the sun and an unknown number of bodies opaque, like our earth; all of which bodies revolve round the sun, and some of which at the same time revolve round others. Those which revolve round the sun only, are called primary planets and comets. Those which revolve round a primary planet, at the same time that they are revolving round the sun, are called secondary planets moons or satellites…. The sun and all the planets, primary and secondary, are globular, though not perfect globes. (Wilkins, 1841)

Kepler (1611) also coined the term servant, guard, or attendant for a planet that orbits another planet, writing the word in Latin as *satellitibus* (dative plural; the nominative plural being *satellitēs*). The transliteration *satellite* entered English from French. The literature shows that astronomers always understood that a satellite, servant, guard, or attendant of a planet was itself also a planet (e.g., Huygens, 1698, Du Séjour, 1789, and Pattillo, 1796), just as a human that is a servant or attendant of another human is still a human; the term *satellite* was always merely occupational for planets.[12] Galileo (1632) coined the category word *moon* after the proper name of the Moon because it was the prototype of planets orbiting planets, just as the category word *geyser* came from the proper name of the prototype Geysir in Iceland. Because Galileo and his audience, geocentrists and heliocentrists alike, understood the Moon to be one of the most important planets and described it as such, they could not think that calling the secondary planets *moons* was a replacement for thinking of them as planets. It simply put them into the subcategory of planets that orbit other planets like the prototype.

[12] In the modern period this is still the case: a star that is a satellite of another star is still a star, and a galaxy that is a satellite of a galaxy is still a galaxy. The idea that a satellite of a planet is *not* a planet came from the 1800s folk taxonomy that we discuss in Section 4.

## 3.3 Later Theoretical Coherence with the Taxonomy

By the early- to mid-1700s, scientists increasingly worked from the perspective that planets form naturally in the cosmos. This created further coherence for a *planet* concept that included both primaries and secondaries because scientists recognized that these subclasses formed out of the same material by the same processes as one another (Kant, 1755; de Laplace, 1795; Trowbridge, 1865; Kirkwood, 1867; Davidson and Stuvé, 1874). We found no arguments in the literature that the formation processes would be essentially different for moons than for primaries or that it would give them any different characteristics, either chemical or geological, than primaries. Instead, we found arguments that supported their commonality. Isaac Newton's nephew, Mr. Conduit, recorded a conversation he had with the 83-year-old Newton in 1724/5 that serves as a very early example of this:

> He [Newton] then repeated to me [Conduit], by way of discourse…that it was his conjecture (he would affirm nothing) that there was a sort of revolution [or recycling of material] in the heavenly bodies; that the vapours and light, emitted by the sun, which had their sediment, as water and other matter, had gathered themselves, by degree, into a body, and attracted more matter from the planets, and at last made a secondary planet (viz: one of those that go round another planet), and then, by gathering to them, and attracting more matter, became a primary planet; and then, by increasing still, became a comet,[13] which after several revolutions, by coming nearer and nearer to the sun, had all its volatile parts condensed…He said he took all the planets to be composed of the same matter with this earth, viz: earth, water and stones &c., but variously concocted. (Chittenden, 1848)

This conversation showed that Newton's *planet* concept was not based on orbital hierarchies; it shows his belief that dynamical states of planets are prone to change; and it shows his belief that the causal difference between the primary and secondary orbital states is the size that resulted from accretion, so moons and primaries exist ontogenetically in a continuum of sizes rather than as fundamentally distinct types of objects.

By 1867, Kirkwood's view on planet formation was recognizably closer to the modern view than Newton's, and it still supported the belief that moons and primaries are the same type of object so he classified them as such. He wrote, “[T]he history of formation…[of every planet], both primary and secondary, would be precisely similar” (Kirkwood, 1867). Scientists also thought that the ring system around Saturn was analogous to the asteroid belt around the Sun (we now know this was incorrect), and they made theoretical arguments about planet formation for the primaries based on observations of the secondaries and vice versa (Peirce, 1851; Kirkwood, 1860; 1869; Proctor, 1865; Trowbridge, 1865). Laden with this theory, the taxonomy continued to be pragmatic keeping moons and primaries together as planets.

[13] At the time, scientists thought comets were larger than ordinary planets. There are many papers that say comets are a type of planet, e.g., "Of Comets" (Anonymous, 1767). We think it is probable that astronomers did not define comets as a subcategory of planets because it was speculative. Indeed, in the conversation recorded by Mr. Conduit here, Newton insisted that he would not conjecture publicly about these things.

Scientists also understood, like Newton, that dynamical states are not permanent, neither during the planetary formation stage nor after it is complete (e.g., Gregory and Halley, 1715; Newton, 1724/5; Trowbridge, 1865; Moulton, 1935; Kuiper, 1951), so primaries can become secondaries and secondaries can become primaries, rogue planets can become primaries and primaries can become rogue planets. As early as 1686, de Fontenelle (1686) discussed how a planet can be spun free from its parent star and enter a neighboring star system, passing through the orbits of the other star’s planets. Therefore, a taxonomy aligned with their present-day dynamical states would not be laden with the most explanatory theory.

These insights have not fundamentally changed in the modern view. We understand moons formed by a variety of processes: sometimes accreting directly as satellites from the original solar disk material (e.g., the Galilean moons), sometimes by capture of a primary body (e.g., Triton), and sometimes through re-accretion following a collision (e.g., the Moon). The observed dynamical hierarchy of Solar System bodies is descriptivist within the timeframe of human perspective but not reductionist since it does not define classes of objects aligned with ontogenetic theory. It also fails, in the modern view, to align the classes of objects with their emergent properties including chemical, geological, and possibly biological. Tidal force is the most likely feature that satellites have in common as a class, but even this feature falls short of defining a pragmatic taxonomy since not all planet-sized satellites have geology dominated by tidal forces with their primary (e.g., Dione), whereas some primary planets have been strongly affected by tidal forces with the Sun (e.g., Mercury), and other bodies have been shaped by tidal forces only before or after changing dynamical states (e.g., Triton). There is no reason to think exoplanets will reduce the statistical diversity of tidal effects among satellites and primaries. More importantly, it is unlikely we could show that the differences caused by tidal effects are more significant than the other differences between planetary bodies derived from the wide variety of factors; tidal effects are just one more factor that can be handled in the taxonomy through subcategorizations.

Newtonian dynamics eliminated another architectural assumption of geocentrism by showing that no stars can be dynamically “fixed”. By the early 1700s, astronomers understood that the so-called fixed stars are simply more massive than the wandering ones, so their motions are relatively smaller:

> [T]he Sun it self is mov’d, and no sensible Body doth rest in the Center; the Center of Gravity of the whole System is chosen for the real Center of our

> World…and the Center of the Sun comes very near to it…There is therefore no perfect Rest of a real Being in Nature. (Whiston, 1716).

Later, the intrinsic difference between stars and planets was also explained as a difference in mass. Mayer hypothesized in 1841 that meteorites falling into the Sun provide its luminous energy, then in the mid-1800s Helmholtz attributed it to gravitational contraction of the Sun's own mass (Kragh, 2016), so producing light is an ordinary property of all celestial bodies if they are large enough. By 1920 Eddington proposed particle physics can sustain luminosity much longer than gravitational contraction, and finally in 1939 Bethe explained it through fusion sustained by adequate mass (Kragh, 2016). Differences in mass thus explained both the dynamical and the luminous differences of stars and planets, putting them onto an ontogenetic continuum.[14]

## 3.4 Universal Agreement Among Scientists

Usage throughout history until the 1920s shows that the community understood planetary satellites to be planets. There are indeed texts where they were just called moons or satellites, so it is impossible from those texts alone to know whether the author considered them to be planets, but the preponderance of data shows there was a recognized uniformity of understanding and an absence of questioning it. Proctor (1880) had to explain to the public that the Moon is a planet, but among scientists they usually took it for granted. There are many cases of ordinary usage in science publications where moons were simply called "planet" without qualification or apology. Textbooks and encyclopedias were written to give an honest representation of the body of knowledge to the non-specialist, so it is telling that they gave formal definitions of *planet* that included moons.[15] An example in 1765 from the influential *Encyclopédie*, the French encyclopedia of the Enlightenment, says,

> SATELLITE in Astronomical terms means secondary planets which move around a primary planet [*des planetes secondaires qui se meuvent au-tour d'une planete premiere*], like the Moon does in relation to the Earth. They are called so because these planets [*ces planetes*] always accompany their primary planet and make their revolution around the Sun with it… The Satellites of Jupiter are four small secondary planets [*quatre petites planetes secondaires*] which revolve around this planet as it itself revolves around the Sun... Galileo, to honor his patron, called these planets [*ces planetes*] the *astra Medicea*… The Satellites of Saturn are five small planets [*cinq petites planetes*] which turn around Saturn ... One of these planets [*Une de ces planetes*], namely the fourth counting from Saturn, was discovered by M. Huygens on March 25, 1655... (d'Alembert, ca. 1765, transl. from French)

Another typical example from an 1870 astronomy manual said,

---

[14] We leave discussion of the modern evolution of the concept of free floating planets, their ontogeny, and their taxonomical relationship to brown dwarfs and bound planets to future work. We think it has been complicated by the historical issues and presentism that we discuss in this paper.

[15] We note in passing that the recent idea that no formal definition of "planet" had been written by astronomers prior to 2006 is incorrect.

> There are two kinds of planets; *Primary* and *Secondary Planets*…PRIMARY PLANETS are those which revolve around the sun only…SECONDARY PLANETS, generally called SATELLITES, are those which revolve around their primaries, and with them, around the sun… There are eight large primary planets in the solar system, besides a great number of smaller ones, called MINOR PLANETS, or ASTEROIDS…The MINOR PLANETS are very small planets which revolve around the sun, between the orbits of Mars and Jupiter. *Ninety-six* have been discovered… (Kiddle, 1870; bold, italics, and small caps in the original]

Fig. 2 is an example of a book (Nasmythe and Carpenter, 1874) that demonstrates how *planet*, *world*, and *satellite* had different but complementary meanings. Different chapters of the book discussed each of these three aspects of the Moon; those discussing its planethood focused on the Moon's intrinsic geology.

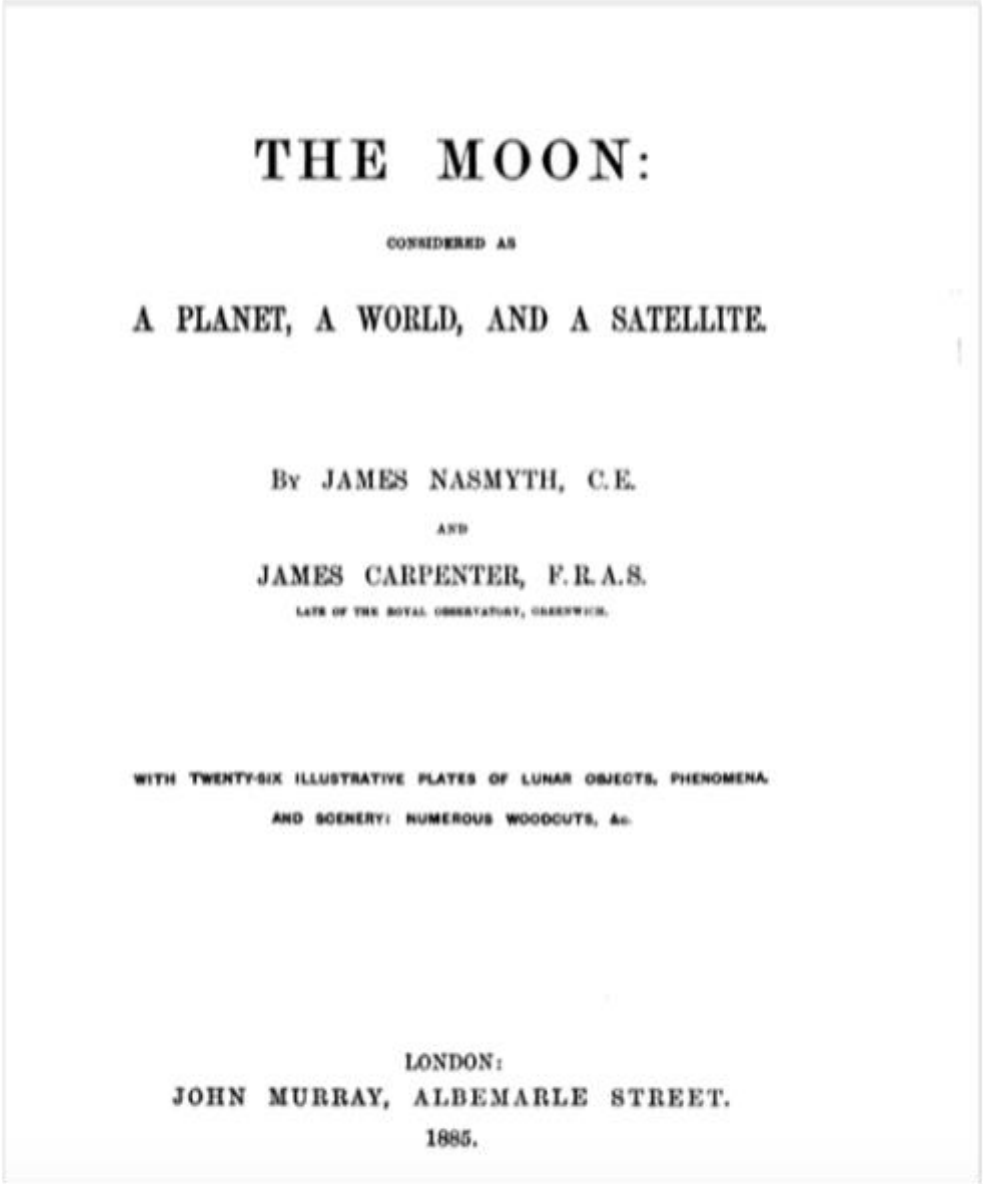

THE MOON:

CONSIDERED AS

A PLANET, A WORLD, AND A SATELLITE.

BY JAMES NASMYTH, C.E.

AND

JAMES CARPENTER, F.R.A.S.

WITH TWENTY-SIX ILLUSTRATIVE PLATES OF LUNAR OBJECTS, PHENOMENA, AND SCENERY; NUMEROUS WOODCUTS, &c.

LONDON:
JOHN MURRAY, ALBEMARLE STREET.
1885.

**Figure 2.** Cover page of an 1874 book considering the Moon "as a planet, a world, and a satellite."

# 4. EMERGENCE OF A FOLK TAXONOMY, CA. 1800 - 1860

Since the Moon was already a planet before the Copernican Revolution, and no effort was made to change that during the Copernican Revolution, and heliocentric scientists uniformly defined planets to include moons after the Copernican Revolution, where then did the idea that moons are not planets come from? As we show in the next section, astronomers abandoned the *planet* concept that had been passed down from the Copernican Revolution sometime after 1920 and switched to some other basis that results in moons being non-planets. If you examine only the scientific literature, that event seems to occur with no theory-based explanation and no historical cause. To find its origin we had to investigate the non-scientific literature. We found that during the 1800s the non-scientific public in the Latin west developed its own folk taxonomy about planets reflecting the concerns of astrology and theology, and that this folk taxonomy eventually affected the scientists. Here we assess (1) *when* this folk taxonomy developed, (2) *why* the public was motivated to develop a folk taxonomy, (3) *how* it developed – i.e., what channels for socializing the folk taxonomy were sufficiently effective that it could prevail over the scientific one – and finally (4) *what* exactly developed: what was the public's theoretical concern that co-evolved with their taxonomy in which moons are not planets? To do this, we scoured the published records over the relevant centuries, selected and studied hundreds of documents, and read the analyses of professional historians.

## 4.1 When It Developed

It is obvious that the modern folk concept could not develop until the public adopted heliocentrism, and that did not occur until the Enlightenment during the mid- to late-1700s. Before that time, the public necessarily retained a geocentric *planet* concept in which the Sun and Moon are planets while the Earth is not. Isaac Watts attested to the general ignorance of Copernicanism as late as 1743, two centuries after Copernicus:

> How much is the vulgar Part of the World surprised at the Talk of the *diurnal and annual Revolutions of the Earth*?....Tell these Persons that the *Sun* is fixed in the Centre, that the *Earth* with all the *Planets* roll round the Sun in their several Periods, and that the *Moon* rolls round the Earth in a lesser Circle, while together with the Earth she is carried round the Sun; they cannot admit a Syllable of this new and strange Doctrine, and they pronounce it utterly contrary to all Sense and Reason….[T]hey look upon these things as Tales and Fancies, and will tell you that the Glasses [telescopes] do but delude your Eyes with vain Images…(Watts, 1743)

Stimson (1917) and Winik (2002) argued that the public embraced Copernican cosmology beginning from the mid $18^{th}$ century with widespread adoption by the end of that century. "Widespread adoption" is a relative term: modern polls regularly show that about 15% to 30% of Europeans, Americans, and Russians still believe geocentrism instead of heliocentrism (Crabtree, 1999; VTsIOM, 2011; Neuman, 2014; Cooper and Farid, 2016). The inability of the educational systems and/or popularizing science communicators of the $17^{th}$ and $18^{th}$ centuries to convince the public suggests just how easy it would be for the public to continue ignoring the scientists' taxonomy even after adopting heliocentrism.

A more specific indicator of when the folk taxonomy developed is found in textbooks written by educated persons who were not professional astronomers. In the first half of the $19^{th}$ century, with almost no exception, those texts were strongly aligned with the Copernican taxonomy of professional scientists, teaching that moons are planets (Patillo, 1796; J. Morse, 1805; Webster,

1807;  J. Morse, 1814; Mather, 1815; Sherwood, 1818; Gomez, 1820; Gregory, 1820; Imison, 1822; Carey, 1825; Taylor, 1825; Vose, 1827; Anon., 1829; Darley, 1830; Blake, 1834; Guy, 1935; Bradford, 1837; Funk, 1837; Anon., 1838; Dick, 1838; Wilkins, 1841; Savile, 1842; Chambers and Chambers, 1843; Cornwell, 1847; Swan, 1848; Frost, 1850; Mattison, 1851; Newton, 1854; Smith, 1856). Presumably, when a non-astronomer wanted to write a book teaching astronomy to the general public, they would consult the writings of professionals, so this is no surprise. Starting about 1857 this type of book underwent a transition, with most publications after that date using the folk taxonomy teaching that only the primaries are planets (McGuffey, 1857; Miller, 1873; Steele, 1874; Giberne, 1880; Sharpless and Philips, 1882; Giberne, 1885; Scovell, 1894; Holden, 1903), although some others continued using the Copernican taxonomy (Tegg, 1858; Anon, 1865; Gavarrete, 1868; Kiddle, 1870; Johnstone, 1887; Leahy, 1910). This indicates a change had occurred by 1857. We give further evidence below that the folk taxonomy began developing by about 1800, growing stronger in the 1820s to 1840s, and becoming clearly preferred by non-astronomers over the alternatives by the 1850s to 1880s. By 1880, English astronomer Richard Anthony Proctor remarked how the public held a different view than the astronomers about the Moon:

> The moon, **commonly regarded** as a mere satellite of the earth is in truth a planet, the least member of that family of five bodies circling within the asteroidal zone, to which astronomers have given the name of the terrestrial planets.
> (Proctor, 1880, bold added)

We also note that the folk taxonomy appeared in the literature precisely when there was a bloating of the planet lists of professional astronomers due to discovery of many satellites and asteroids. Since the folk taxonomy had the effect of simplifying the list of planets, making it more like the old geocentric list, the correlation of that timing seems significant.

## 4.2 Why It Developed

It should be noted that a concept like *planet* is not held in our minds as a definition of the sort provided by the IAU in 2006. Cognitive scientists and philosophers of science argue that a concept is a complex of relationships that may include composition, setting, function, purpose, and more (Thagard 1990b; Giere 1994). The goal here is to determine the key elements of the public’s concept for planets in the 1800s, what the public thought was important about planets in lieu of, or in addition to, the intrinsic geological properties favored by Copernican scientists. We discovered and investigated three significant trends in the literature. We found one of these to be uncorrelated to the appearance of the folk taxonomy while the other two appear to be causal for its development.

### *4.2.1 Questions of Habitability*

When scientists rejected Aristotelianism and accepted the geophysical/geological *planet* concept forged by Galileo (Fabbri, 2012; 2016), they immediately began making teleological arguments that the planets must be inhabited: if planets are other earths, then like the Earth they must be the homes of other worlds or else why should they exist? (Kepler, 1610; Wilkins, 1638; de Fontenelle, 1686; Huygens, 1695; Brewster, 1854). See the analyses by Dick (1984), Crowe (1988), and Crowe and Dowd (2013). Amidst their other teleological arguments, some of the scientists suggested distinct teleological roles for the primaries versus the secondaries regarding habitability. Kepler (1610) started this by arguing that the moons of Jupiter serve the world of intelligent creatures on the primary by providing them with light, giving the moons purpose apart from being habitable worlds themselves. Much later, Davis (1868) stated that the primaries *as a class* are the ones that are inhabitable. Newcomb and Holden (1880) noticed that the geophysical similarity that led Galileo to put the Earth and the Moon into the same class had been weakened as better science showed the Moon to be uninhabitable. We investigated whether this distinction motivated the emergence of the folk taxonomy making satellites out to be non-planets. We found that there was no correlation. During the period in question other scientists were arguing that the secondaries are inhabited (e.g., Bode, 1801). One amateur astronomer even reversed the roles, arguing that the primary Jupiter serves the worlds that exist on its secondaries by giving them warmth and light to enable habitation, so the secondaries would have the purpose of being worlds while their primary Jupiter would have the subservient purpose (Giberne, 1880). We found that those who argued the secondaries are uninhabitable still considered them to be planets, and there was no trend with any view taking ascendancy over the others while the folk taxonomy developed. Those who were most attuned to the differences of habitability were not members of the general public where the folk taxonomy was developing, but scientists who were the ones resisting the folk taxonomy. Scientists still held that moons were planets long after the teleological arguments had been discarded in the later 1800s (Crowe, 2016). Apparently, the theories of planet formation and emergent geology produced adequate coherence between moons and primaries to continue supporting the taxonomy, so teleological similarity was no longer required. The questions of habitability, apart from teleology, continued unbroken into the present and remained central to planetary science and it is still recognized that both primaries and secondaries can be locations for life. Our literature review showed conclusively that the question of habitability was not what motived the emergence of the folk taxonomy.

#### *4.2.2 Trend from Astrology: Planets Must Be Simple*

The literature demonstrates that during the relevant period there was ongoing, strong interest in astrology. Historians have shown that during the 18th century, occult beliefs in general were pushed out of the upper echelons of society in both Europe and the Americas but they survived in popular culture. Historians call that the *folklorization* of the occult. Butler (1992) wrote,

> The persistence of belief in witches after the trials had ended reflected the folklorization of magic in the twilight of early modern Western society on both sides of the Atlantic. Although upper social classes largely abandoned occultism, other colonists continued to believe in witchcraft, **astrology**, and the ability of wise men and wise women to find lost objects and cure disease. In this regard, folklorization prevented the complete suppression of occultism and magic. (Bold added)

Winik (2002) wrote,

The result of this intellectual quagmire was the survival of occult astrology through the eighteenth century as a *popular* force, immune to the onslaught of scientific and intellectual elites. [italics in the original]

The public's old geocentric and emerging heliocentric *planet* concepts rode along in the folklorization of astrology. Astrology depends on a crucial assumption about the planets: they must be few enough and orderly enough that their influences can be sorted out and turned into predictions. Belief in planetary influences was a worldwide phenomenon, so this assumption must have evolved through the limited perspective of humanity without telescopes, as only a few lights could be seen moving in orderly ways about the sky. Since people in the 18$^{th}$ and 19$^{th}$ centuries still wanted to believe astrology after switching to heliocentrism, the logic of astrology must have imposed a subconscious bias toward continuing to see planets in the way that is necessary for astrology to make sense. That is, it must have imposed a bias toward seeing planets as a small and orderly set of objects.

By the latter half of the seventeenth century, the astrologers had transitioned their livelihoods from serving wealthy patrons to selling almanacs, which became extremely popular among the general public (Dohoney, 2011). They were a mainstay of popular culture throughout the eighteenth century (Dragoş, 2018) and a third of a million were printed annually in England (Capp, 1979). It is estimated that in America by the 1800s, enough almanacs were sold yearly for every household to obtain a copy (Tomlin, 2010). An example of astrological prognostication sold to the public in 1795, a century after the intellectuals had abandoned it, is shown in Fig. 3.

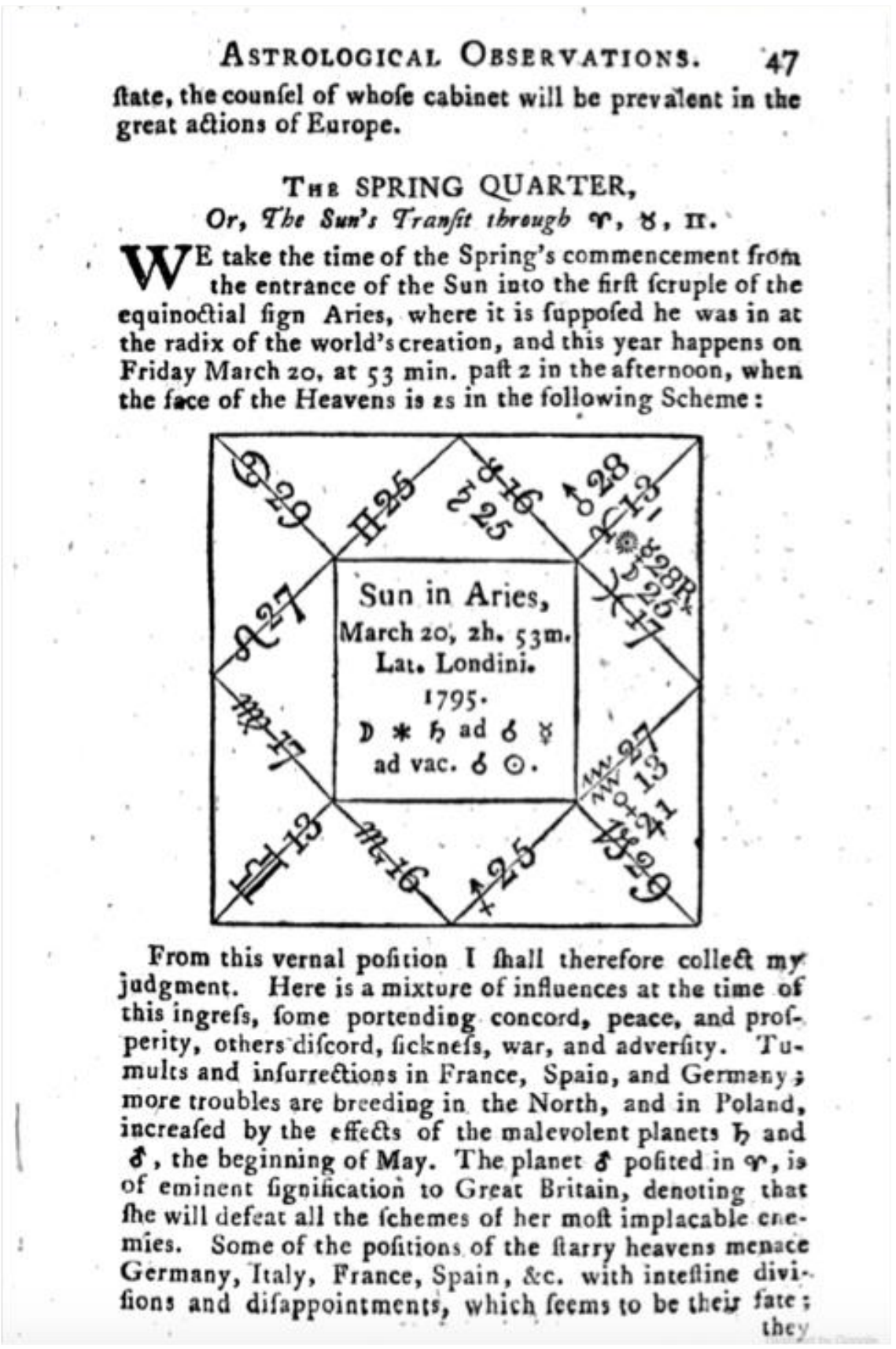

ASTROLOGICAL OBSERVATIONS. 47

ftate, the counfel of whofe cabinet will be prevalent in the great actions of Europe.

THE SPRING QUARTER,
*Or, The Sun's Tranfit through* ♈, ♉, ♊.

WE take the time of the Spring's commencement from the entrance of the Sun into the firft fcruple of the equinoctial fign Aries, where it is fuppofed he was in at the radix of the world's creation, and this year happens on Friday March 20, at 53 min. paft 2 in the afternoon, when the face of the Heavens is as in the following Scheme:

From this vernal pofition I fhall therefore collect my judgment. Here is a mixture of influences at the time of this ingrefs, fome portending concord, peace, and profperity, others difcord, ficknefs, war, and adverfity. Tumults and infurrections in France, Spain, and Germany; more troubles are breeding in the North, and in Poland, increafed by the effects of the malevolent planets ♄ and ♂, the beginning of May. The planet ♂ pofited in ♈, is of eminent fignification to Great Britain, denoting that fhe will defeat all the fchemes of her moft implacable enemies. Some of the pofitions of the ftarry heavens menace Germany, Italy, France, Spain, &c. with inteftine divifions and difappointments, which feems to be their fate; they

**Figure 3.** A page on astrology from a London almanac in 1795 (Partridge, 1795).

The list of planets taught in these almanacs during the 18th and early 19th centuries varied widely according to the preferences of their authors, but they all displayed the basic assumption of simplicity and orderliness inherent to astrology. Most of the public never received any formal education about the planets (see below), so these almanac planet lists were the main, or the only, source of information about planets that most members of the public ever received, and they shaped the *planet* concept that the public was learning. Despite the story told in modern astronomy texts about how the concept of *planet* changed with Copernicus, there was in fact no law that said the Earth had to become a planet because it moves and that the Sun had to stop being a planet because it is a fixed star. Popular culture could, and did, develop its own logic. We see that reflected in these lists.

Three main families of planet lists are apparent during this period. The first of these families was essentially geocentric. Even after switching to heliocentrism, many almanacs continued to give lists that were seven planets including the Sun and Moon but not the Earth. After its discovery, Uranus[16] was readily added to these lists, making eight planets, and eventually Uranus was supplied its own astrological interpretation. It would have been inconsistent for there to be a planet without astrological influence. Three examples from the 1800s of this very common family of lists are shown in Fig. 4.

The second family of lists that appeared in almanacs are the ones that added the Earth into an otherwise geocentric set of planets that still included the Sun and the Moon. The Earth was never visible in the sky in a particular part of the zodiac, of course, so it had no function in these almanacs and the symbol for the Earth did not appear anywhere in the almanac's calendrical tables. It was added to the planet lists as a theoretical addition to teach or to acknowledge that heliocentrism is the correct cosmology, not a practical addition for the purposes of the almanac. The Sun, on the other hand, was moving in the sky from our perspective and was needed for the astrology so following the long tradition it was kept on most planet lists and was still explicitly called a planet well into the 1800s. These lists that included both the Earth and the Sun as planets represent a hybridization of theoretical heliocentrism with practical geocentrism. They were common during the 19th century, and four examples are shown in Fig. 5.

The third family of lists in the almanacs avoided calling the Sun a planet or explicitly said that it is not a planet while still including it in the astrological tables. Some examples how this was done are shown in Fig. 6. Some of these almanacs also set the Moon aside as a "satellite" separate from the "planets". These are the versions of the planet lists that are familiar today, not because they are inherently right according to the story told in modern astronomy textbooks, but simply because this is the family of lists that eventually prevailed. The question remains *why* this particular family of lists prevailed.

[16] Uranus was usually called Herschel and sometimes Georgian or Georgium Sidus in the almanacs of this era.

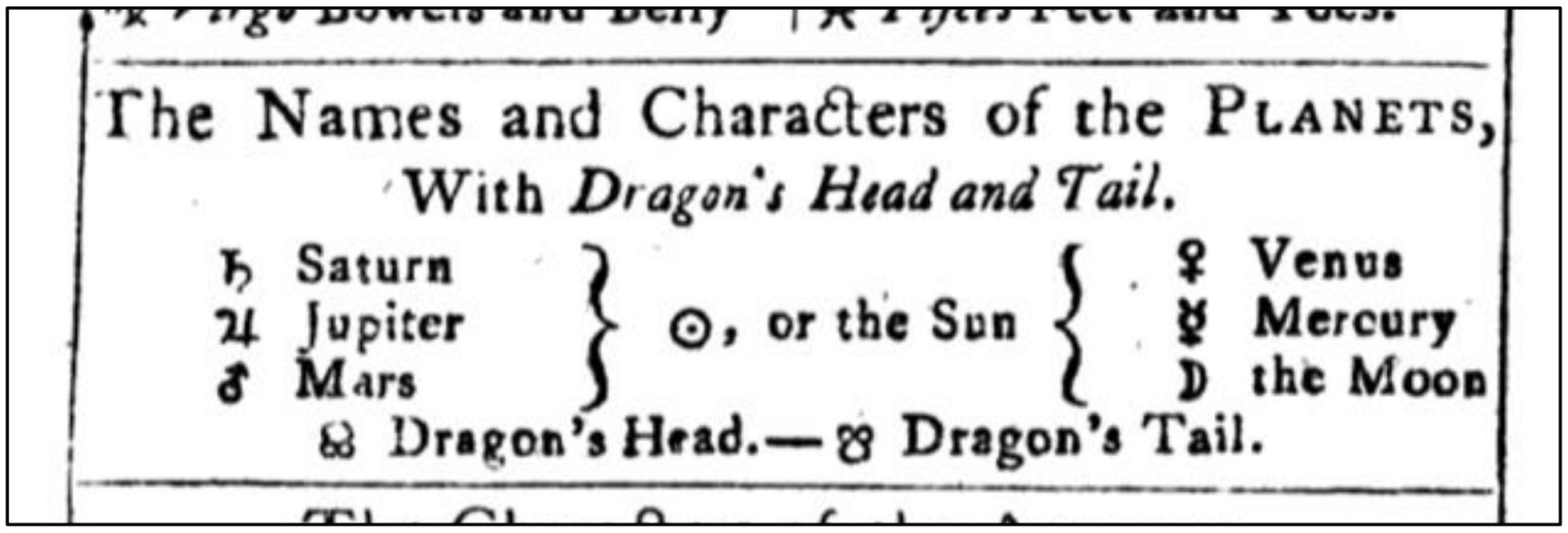
The Names and Characters of the PLANETS,
With *Dragon's Head and Tail.*

♄ Saturn, ♃ Jupiter, ♂ Mars, ☉, or the Sun, ♀ Venus, ☿ Mercury, ☽ the Moon

☊ Dragon's Head.—☋ Dragon's Tail.

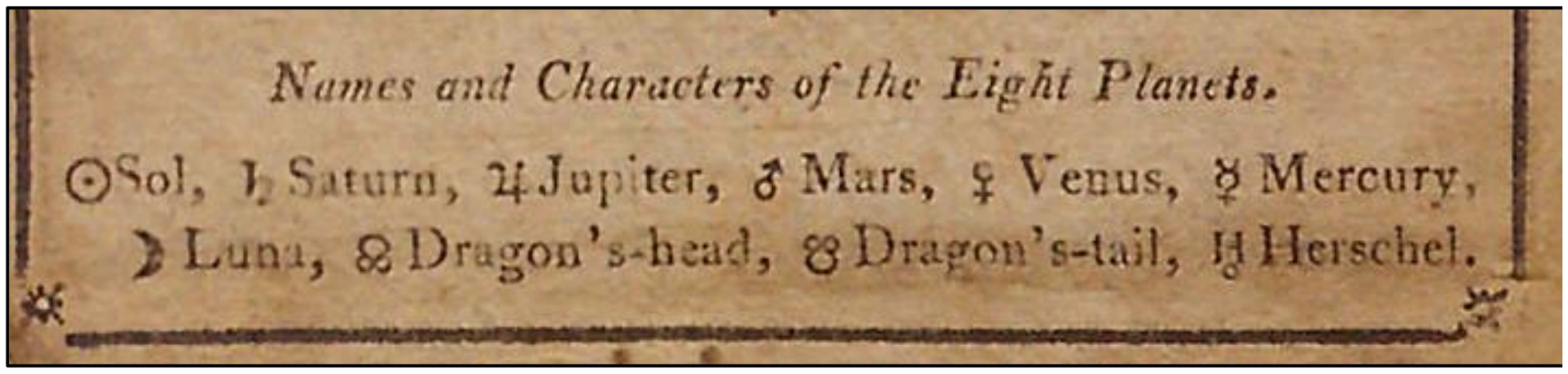
*Names and Characters of the Eight Planets.*

☉Sol, ♄ Saturn, ♃ Jupiter, ♂ Mars, ♀ Venus, ☿ Mercury, ☽ Luna, ☊ Dragon's-head, ☋ Dragon's-tail, ♅ Herschel.

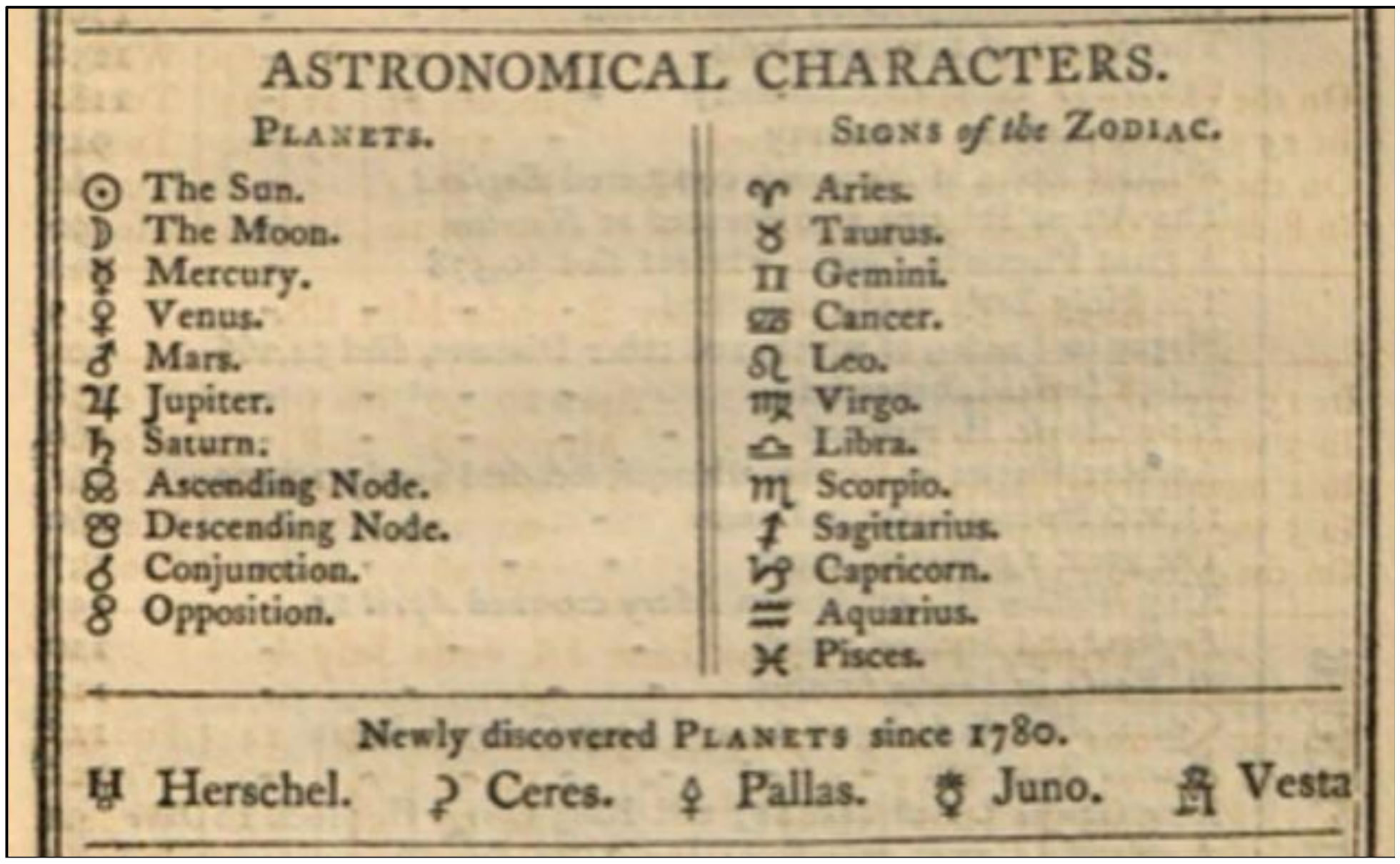
ASTRONOMICAL CHARACTERS.

| PLANETS. | SIGNS *of the* ZODIAC. |
| --- | --- |
| ☉ The Sun. | ♈ Aries. |
| ☽ The Moon. | ♉ Taurus. |
| ☿ Mercury. | ♊ Gemini. |
| ♀ Venus. | ♋ Cancer. |
| ♂ Mars. | ♌ Leo. |
| ♃ Jupiter. | ♍ Virgo. |
| ♄ Saturn. | ♎ Libra. |
| ☊ Ascending Node. | ♏ Scorpio. |
| ☋ Descending Node. | ♐ Sagittarius. |
| ☌ Conjunction. | ♑ Capricorn. |
| ☍ Opposition. | ♒ Aquarius. |
| | ♓ Pisces. |

Newly discovered PLANETS since 1780.

♅ Herschel. ⚳ Ceres. ⚴ Pallas. ⚵ Juno. ⚶ Vesta

**Figure 4.** Example of the (essentially) geocentric planet lists found in popular almanacs in the 19th century. Top: from Moore (1803). Middle: With the addition of Uranus (Herschel), from Keatinge (1806). Bottom: With the addition of Uranus (Herschel) and the first four asteroids, from Poor Robin (1827).

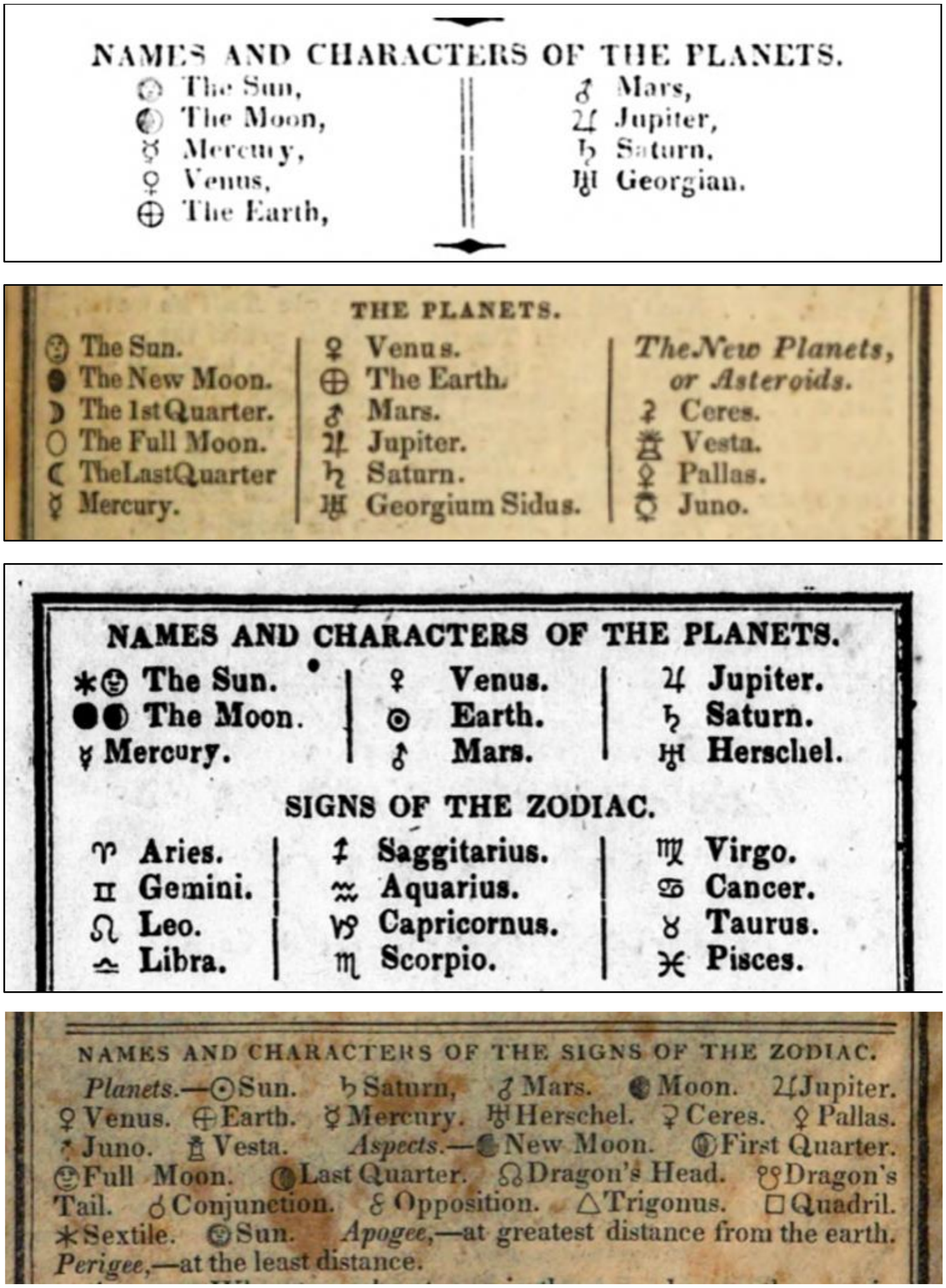

NAMES AND CHARACTERS OF THE PLANETS.

☉ The Sun, ☽ The Moon, ☿ Mercury, ♀ Venus, ⊕ The Earth, ♂ Mars, ♃ Jupiter, ♄ Saturn, ♅ Georgian.

THE PLANETS.

The Sun. The New Moon. The 1st Quarter. The Full Moon. The Last Quarter. ☿ Mercury. ♀ Venus. ⊕ The Earth. ♂ Mars. ♃ Jupiter. ♄ Saturn. ♅ Georgium Sidus. *The New Planets, or Asteroids.* ⚳ Ceres. ⚶ Vesta. ⚴ Pallas. ⚵ Juno.

NAMES AND CHARACTERS OF THE PLANETS.

☉ The Sun. The Moon. ☿ Mercury. ♀ Venus. ⊙ Earth. ♂ Mars. ♃ Jupiter. ♄ Saturn. ♅ Herschel.

SIGNS OF THE ZODIAC.

♈ Aries. ♊ Gemini. ♌ Leo. ♎ Libra. ♐ Saggitarius. ♒ Aquarius. ♑ Capricornus. ♏ Scorpio. ♍ Virgo. ♋ Cancer. ♉ Taurus. ♓ Pisces.

NAMES AND CHARACTERS OF THE SIGNS OF THE ZODIAC.

*Planets.*—☉Sun. ♄Saturn. ♂Mars. Moon. ♃Jupiter. ♀Venus. ⊕Earth. ☿Mercury. ♅Herschel. ⚳Ceres. ⚴Pallas. ⚵Juno. ⚶Vesta. *Aspects.*—New Moon. First Quarter. Full Moon. Last Quarter. ☊Dragon's Head. ☋Dragon's Tail. ☌Conjunction. ☍Opposition. △Trigonus. □Quadril. ✱Sextile. Sun. *Apogee,*—at greatest distance from the earth. *Perigee,*—at the least distance.

**Figure 5.** Example hybrid geocentric/heliocentric planet lists from the 19th century including both the Sun and the Earth as planets. Some include asteroids and others do not. From top to bottom: Anonymous (1831); Anonymous (1834); Anonymous (1836); Anonymous (1839).

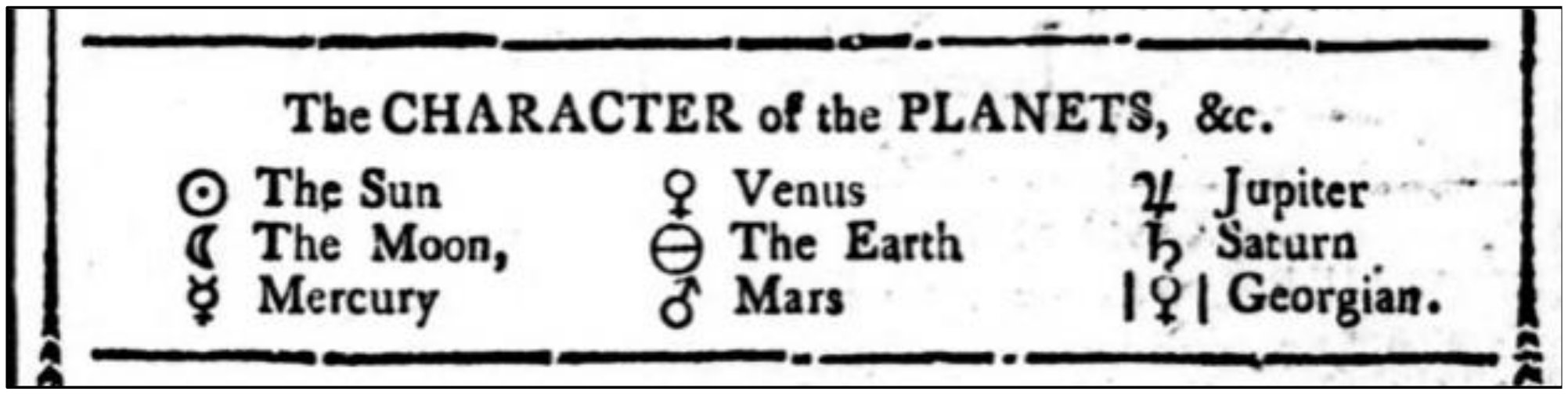

The CHARACTER of the PLANETS, &c.

| | | |
|---|---|---|
| ☉ The Sun | ♀ Venus | ♃ Jupiter |
| ☾ The Moon, | ⊖ The Earth | ♄ Saturn |
| ☿ Mercury | ♂ Mars | ♅ Georgian. |

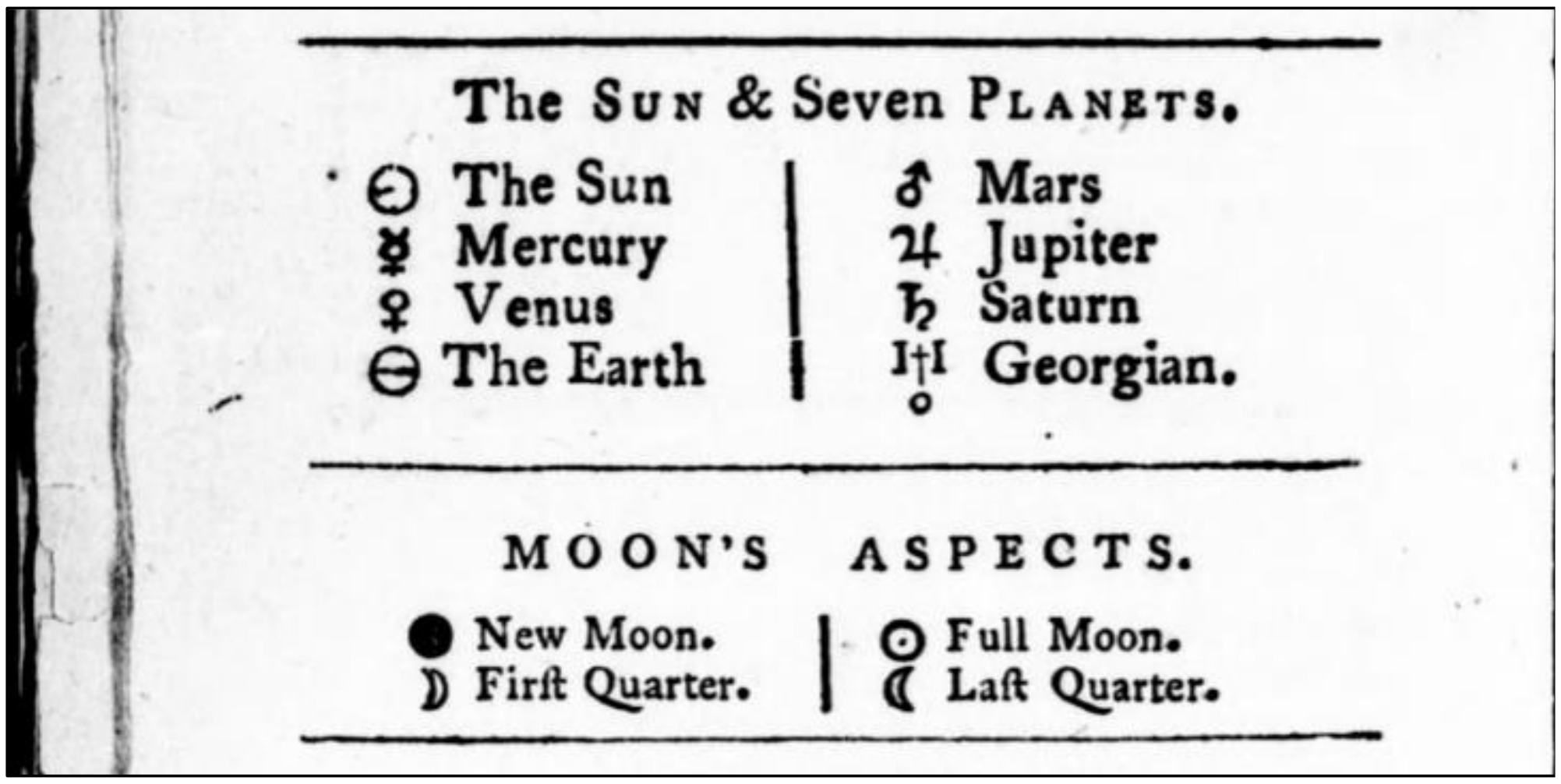

The Sun & Seven Planets.

| | |
|---|---|
| ☉ The Sun | ♂ Mars |
| ☿ Mercury | ♃ Jupiter |
| ♀ Venus | ♄ Saturn |
| ⊖ The Earth | ♅ Georgian. |

MOON'S ASPECTS.

| | |
|---|---|
| ● New Moon. | ☉ Full Moon. |
| ☽ Firſt Quarter. | ☾ Laſt Quarter. |

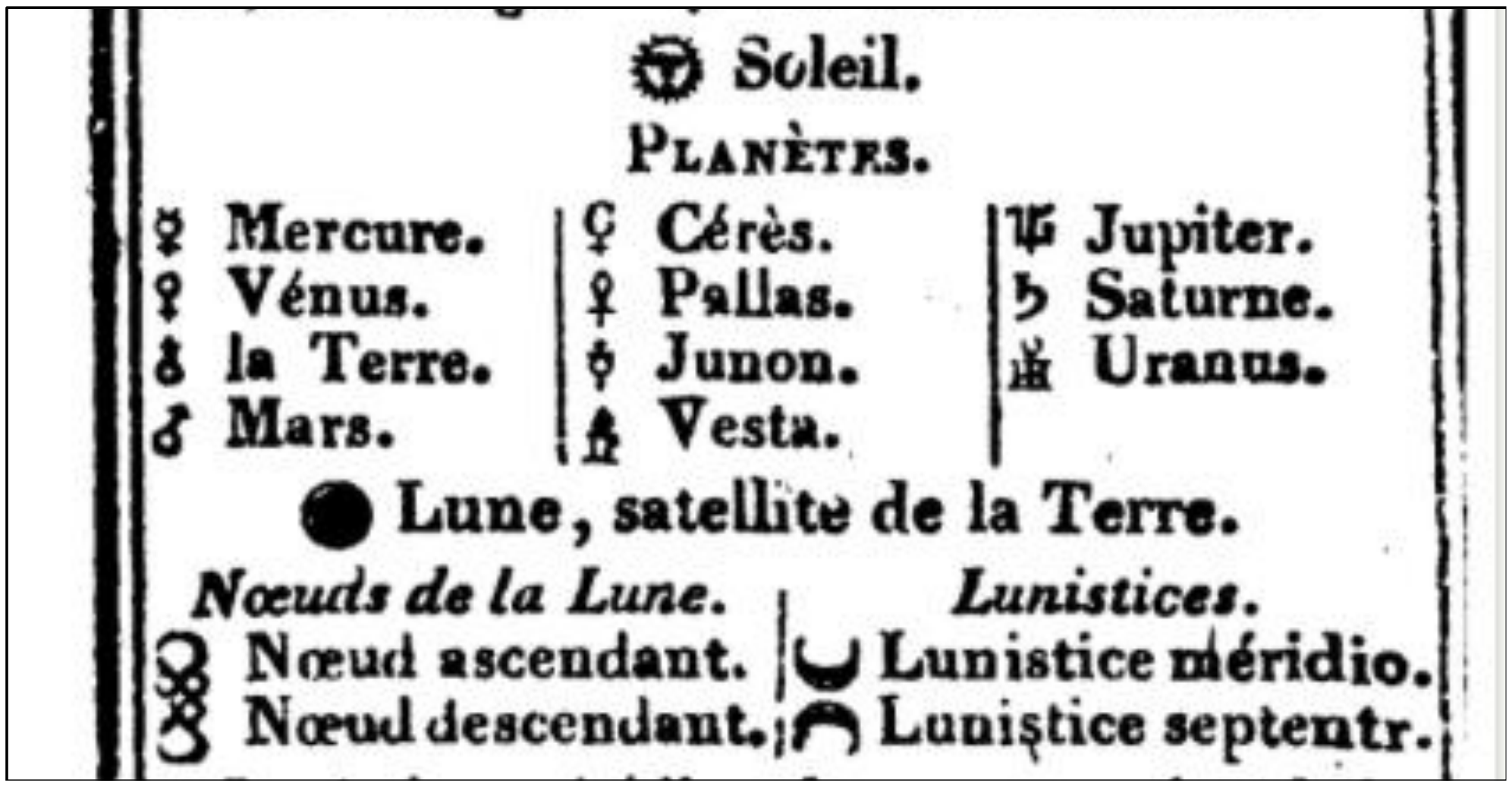

☉ Soleil.

Planètes.

| | | |
|---|---|---|
| ☿ Mercure. | ⚳ Cérès. | ♃ Jupiter. |
| ♀ Vénus. | ⚴ Pallas. | ♄ Saturne. |
| ♁ la Terre. | ⚵ Junon. | ♅ Uranus. |
| ♂ Mars. | ⚶ Vesta. | |

● Lune, satellite de la Terre.

| *Nœuds de la Lune.* | *Lunistices.* |
|---|---|
| ☊ Nœud ascendant. | Lunistice méridio. |
| ☋ Nœud descendant. | Lunistice septentr. |

**Figure 6.** Examples of planet lists that avoid calling the Sun and/or the Moon a planet. Top: simply adding "&c" to the title to acknowledge not everything in the list is a planet (Metonicus, 1799). Middle: Distinguishing the Sun from "Seven Planets" (Anonymous, 1800). Bottom: Placing the Sun and the Moon outside the table (*L'Astrologue Normand*, 1837).

All three families of planet lists were extremely simple compared to the planet lists that professional astronomers were using. By 1789 there were 14 known secondary planets including Earth's Moon, and by 1851 there were 18 secondary planets plus 33 minor planets or asteroids. By 1860 the number of asteroids was 62. Scientists were listing all of these as planets (Cornwall, 1847; Smith, 1856; Kiddle, 1870; Gore, 1893). On the other hand, there were just seven bodies used in astrology, or eight after Uranus was invested with astrological meaning. Many almanacs added the first four asteroids to the planet tables but never found any astrological meaning for them, and we found one almanac that listed the first ten (Athanæum, 1849), and as an exception to the rule an Italian almanac listed forty (Anonymous, 1857), but the vast majority listed either zero asteroids or just four even after dozens had been discovered. As Dohoney (2011) pointed out, astronomy was discovering too many planets to use in astrological prognostication. Discovering an interpretation for Uranus without the voice of antiquity to give it credibility was a trick but inventing astrological meanings for literally dozens of new planets was not feasible.

Furthermore, the secondary planets in the outer solar system are always in the same parts of the zodiac as their primaries, so by definition it would be impossible for their orbital positions to send a distinct astrological "signal" to affect the Earth, at least to the level of accuracy that astrology customarily measured these things. Therefore, it would have been meaningless or self-defeating to include them among the planets.

The family of planet lists that omitted Earth had become less common after about 1810. It would have been possible, in principle, to keep updating these lists to include more asteroids, except that the sheer numbers of non-functional planets would have undermined the basic assumptions of astrology, proving the cosmos to be less orderly and less decipherable than astrologers had always claimed it to be. Including only the Earth's satellite as a planet but not all the other satellites was another inconsistency; although the Moon was the only satellite presumed capable of sending an astrological "signal" to the Earth, it made the lists of planets incomplete from an objective perspective. There are a surprising number of 18th and 19th century popularizing works that say, "the Moon is not a primary planet" or "the Moon is not a planet" (the context usually indicating it meant primary) (Keill, 1730; Ferguson, 1757; Burges, 1789; Walker, 1790; Walker, 1802; Prändel, 1813; Carey, 1825; Newton, 1854). This implies that the public, having recently abandoned geocentrism, needed this clarification. Some almanacs dealt with the inconsistency by taking not just the Sun but also the Moon out of the planet lists and treating them both as special cases, bodies that are relevant to astrology but not actually "planets". In summary, astronomy was proving astrology to be increasingly disconnected from reality, but the almanacs could maintain the fiction of astrological relevance and maintain their sales by selectively manipulating the list of planets they presented to the public.

#### *4.2.3 Trend from Theology: Planets Must Be Orderly*

The second way the public's *planet* concept was informed was through a theological argument that the purpose of planets was to impress on humans the mathematical and symbolic orderliness of the cosmos, both as an indicator of the divine and as an inducement to moral improvement (Gronim, 1999). We found many sources evidencing this type of thought through the 1700s and 1800s (Baker, 1746; Clarke, 1749; Cornwall, 1823; Rowson, 1804; Hervey, 1808; Chalmers,

1818; Godwin, 1831; Turner, 1832; Pierpoint, 1836; Harrowar, 1838; Dick, 1838; Schmucker, 1839; Matthews, 1843; Greenwood, 1847; Hale, 1848; Peck, 1848; Bushnell, 1849; Wordsworth, 1849; Parker, 1853; Carpenter, 1840; Lorrain, 1852; Riddle, 1852; Douglas, 1854; Walter, 1854; McGuffey, 1857; Beer, 1870).

Astrologers were one of the groups interested in promoting the theological argument. Tomlin (2010) argued that they wanted to distance astrology from the occult and make it acceptable to Protestant Christian readers since their livelihoods depended on it. The popular almanacs of the middle 18th century "disseminated a view of the natural world infused with religious significance. Readers were encouraged to pay close attention to its workings, which provided another sacred, revealed, and legible text—made even more immediately legible by the almanac" (Tomlin, 2010).

However, it was not primarily the astrologers who championed this view. We found it mainly in religious books and in sermons that were preached during the era. Planetary orbits were used as a primary datum for natural theology (e.g., Clark, 1749; Prescott, 1803; Hervey, 1808; Bentley, 1809; White, 1811; Price, 1816; Parish, 1826; Smith, 1842; Peck, 1848; Wayland, 1849; Forbes, 1854; Parker, 1856), and they were used in sermons as analogies to illustrate moral attributes such as faithfulness and humility (e.g., Griffin, 1808; Bushnell, 1849; McIlvaine, 1855). Both applications depended on the planets being orderly idealizations reminiscent of the old geocentric system rather than the evolving, chaotic, contingent, geologically complex objects of scientific understanding.

An English clergyman's meditations written in 1746 and republished in 1808 said,

> Yet none [of the planets] mistake their way, or wander from the goal; though they pass through trackless and unbounded fields. None fly off from their orbits into extravagant excursions; none press in upon their centre with too near an approach. None interfere with each other in their perennial passage, or intercept the kindly communications of another's influence. But all their rotations proceed in eternal harmony; keeping such time, and observing such laws as are most exquisitely adapted to the perfections of the whole… How soon, and how easily, is the most finished piece of human machinery disconcerted! But all the celestial movements are so nicely adjusted, all their operations so critically proportioned, and their mutual dependencies so strongly connected, that they prolong their beneficial courses throughout all ages…
>
> While all the stars, that round her burn,
> And all the planets, in their turn,
> Confirm the tidings as they roll,
> And spread the truth from pole to pole. (Hervey, 1808)

Sharon Turner, in a religious work, wrote in 1832 (when only four asteroids were known) about the teleological value of Bode's Law and how that impresses orderliness upon our human minds:

> The asteroids, or telescopic planets, that revolve between Mars and Jupiter need not be further noticed here than to mention their apparent confirmation of the new law which the scientific Bode had suggested... [T]his law seemed to be interrupted between Mars and Jupiter. Hence, he inferred, that there was a planet wanting in that interval; a bold yet profound conjecture; but this predicted deficiency is now found to be supplied by the four new asteroids, which occur in the very space where the unexplained vacancy presented a strong objection to the theory... This establishment of such a law furnishes another impressive instance of the scientific plan and principles on which creation has been fabricated. Every new perception of the intelligent laws by which the heavenly bodies move and are regulated makes more palpable the impossibility that they can have occurred from any other origin than that of a designing, conceiving, selecting, and ordaining cause – a real, pre-existing, intellectual Creator. Such wonderful science, so exactly, so efficaciously, and so permanently operating, can never have arisen from mere confusion, from random motivity, or from irrational chance. (Turner, 1832)

The pattern of orderliness they saw in Bode’s Law included, by definition, only the primaries. Others were looking for separate versions of Bode’s Law within each system of secondary planets (Challis, 1830).

Taking this further, Thomas Dick in 1838 addressed the apparent *disorderliness* of these new asteroids’ orbits. He acknowledged that preserving the Aristotelian view of perfect, unchanging planets was no longer the concern:

> As we know that changes have taken place in our sublunary region since our globe first came from the hands of its Creator, so it is not contrary either to reason or observation to suppose that changes and revolutions, even on an ample scale, may take place among the celestial orbs. We have no reason to believe in the "incorruptibility” of the heavenly orbs, as the ancients [the Aristotelians] imagined, for the planets are demonstrated to be opaque globes as well as the earth; they are diversified with mountains and vales, and, in all probability, the materials which compose their surfaces and interior are not very different from the substances which constitute the component parts of the earth.

However, the heavens must still have been orderly in their original creation:

> Having been accustomed to survey the planetary system as a scene of proportion, harmony, and order, we can scarcely admit that these bodies [the asteroids] move in the same paths, and are arranged in the same order as when the system was originally constructed by its Omnipotent Contriver…The hypothesis of the bursting of a large planet between Mars and Jupiter accounts in a great measure, if not entirely, for the anomalies and apparent irregularities which have been observed in the system of the new planets; and if this supposition be not admitted, we cannot account, on any principle yet discovered, for the singular phenomena which these planets exhibit. (Dick, 1838)

The planets themselves could be messy and geological on their surfaces, and they can even burst apart completely, but their orbits in their original creation must have been well-spaced, non-intersecting, and mathematically harmonious. The concern was specifically with their orbits because that was the one thing that still conveyed order to our minds now that corruptible earthly physics has been admitted onto their surfaces.

The English poet Sara Hale (1848) took this idea further, connecting the asteroids to the origin of human sin. Her poem celebrated only the primary planets – "All the Planets" she called the primaries – celebrating them one-by-one for orderliness and beauty.

> And all these fair Planets in harmony move
> Round the Sun, as their centre of light, life and love.

Where the asteroids are now found she placed "the Guardian Planet", "the largest and loveliest" of the planets, which watched over Earth and in particular guarded the woman in the Garden of Eden, until it was broken into asteroids at the very moment the woman ate the forbidden fruit. Henry Walker (1854) suggested the same idea, writing "if there has been a disruption of a planetary world" forming the asteroids, it cannot be a reflection of their original, perfect estate but "should be respectfully regarded as the result of some moral cause."

William Godwin in 1831 gave us another example of the common idea of orderliness:

> Till Herschel's time we were content with six planets and the sun, making up the cabalistical number seven. He added another [Uranus]. But these four new ones [the first four asteroids] entirely derange the scheme. The astronomers have not yet had the opportunity to digest them into their places, and form new worlds of them. This is all unpleasant. (Godwin, 1831)

Godwin's "six planets" were the primaries, omitting the Moon and the other secondary planets known by that time.

One hundred fifty years earlier, de Fontenelle had also represented his non-scientific companion as shocked to hear a scientist say that a planet can go around another planet because it did not match the expected order, informed as it was by geocentrism:

> "But," interrupted the Marchioness, "how can there be planets that revolve around other planets no better than themselves? Seriously, to me it would seem more regular and more uniform if all the planets, both the large and the small, had only the same sort of movement around the Sun." (de Fontenelle, 1686, transl. from Fr.)

With the simplified heliocentric list of planets, almanac compilers could still signal allegiance to heliocentric reality but do it in a way that makes astrology look plausible, because the planets that send astrological "signals" to Earth are simple enough to convey order and meaning. Even as the popularity of astrology waned during the mid-to-late 19$^{th}$ century, the argument about the divine orderliness of planets was able to survive. Science popularizers sometimes combined this idea with the argument about the plurality of worlds to explain the purpose of planets (e.g., Dick, 1838). Thus, both for their purported astrological effects on Earth and for their direct impressions of the divine, the planets were reshaped to imply cosmological orderliness and meaning like what had existed in geocentrism.

## 4.3 How It Developed

### *4.3.1 Socialization Channels*

The next question is whether these motivations are adequate to explain how the folk taxonomy spread through culture widely enough to become the shared understanding of what planets are, even while scientists were teaching something different. To answer this, we must know something about the channels of socialization. The published record shows there were actually three planet concepts competing for acceptance in the early 1800s: (1) the old geocentric concept, (2) the heliocentric geophysical/geological concept favored by Copernican scientists, and (3) the heliocentric "orderly cosmos" concept that eventually became the common folk taxonomy. The orderly cosmos concept appeared in different forms in the astrological planet lists until one of the versions emerged as the winner. Socialization channels for these competing concepts may be divided into four tiers. The top or "professional" tier socialized ideas between intellectuals, and it included the science journals, letters between scientists, and college-level textbooks written by professional astronomers. The second or "educational" tier included books written by educated non-astronomers, such as professional educators and geographers, specifically intended to teach science about the planets to the public or to students at the primary or secondary levels. The third or "cultural" tier consisted of publications aimed at the public with cultural goals other than to teach science as an end it itself. It included the yearly astrological almanacs, religious writings, sermons that discussed planets, and allusions to the planets in poetry and stories. The fourth tier consisted of oral tradition. A diagram representing these four tiers is given in Fig. 7.

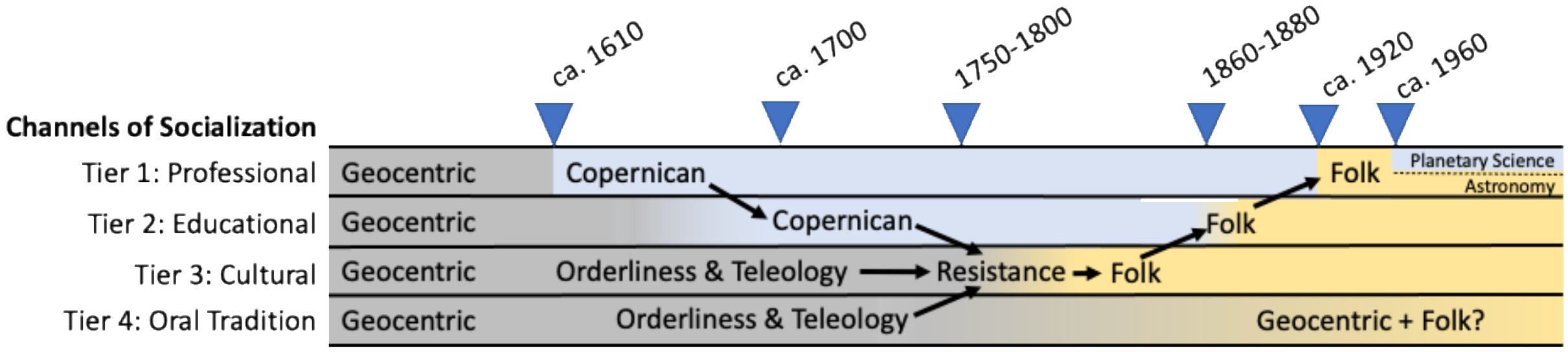

**Figure 7**. Channels of Socialization of the planet concepts showing the dominant *planet* concept found in each channel along the timeline. The period of resistance in Tier 3 publications coincides with the bloating of the Copernican planet lists by satellites and asteroids. Cultural publications resisted this bloat because it upset the orderliness that was necessary in the astrological and teleological interpretations of the cosmos. That led Tier 3 authors to develop the simplified and orderly folk taxonomy in the early- to mid-1800s. We see the folk taxonomy adopted into Tier 2 publications starting 1857 through 1880, then into Tier 1

> publications by the 1920s. Planetary scientists rediscovered and began using the reductionist Copernican taxonomy around 1960, leading to the split between astronomers and planetary scientists when the IAU formally adopted the folk taxonomy in 2006.

In section 3 regarding the scientists' *planet* concept, we discussed the Tier 1 publications. Tier 1 publications socialized only the geophysical/geological Copernican concept during the relevant period. In subsection 4.2 we referenced the Tier 2 textbooks beginning to teach the folk concept ca. 1857. While those texts demonstrate the change, they do not explain why the change occurred. It cannot be explained by a change in the Tier 1 documents, because there was no change in Tier 1 during the relevant period.

Any spreading of the folk taxonomy to explain why the Tier 2 publications changed therefore must have occurred prior to 1857 and must be found in Tier 3 and/or Tier 4. Under "Why It Developed" we cited the Tier 3 publications, and this is where we found motives for the folk taxonomy to develop. We identified a period of development in the almanacs with the correct timing as the public moved generally from geocentrism ca. 1800 to the folk taxonomy by the mid-1800s. We also found many theological writings motivating the folk taxonomy during the relevant period.

Analysis of Tier 4 oral tradition is beyond the scope of this study. Harland and Wilkinson (1882) documented the folklore of Lancashire, England from the mid-1800s and reported deeply ingrained superstitions related to astrology and the planets, and many people still believe geocentrism to this day (Crabtree, 1999; VTsIOM, 2011; Neuman, 2014; Cooper and Farid, 2016). From these sparse indicators, we can reasonably guess that geocentrism and astrology were among the influences in the early 1800s in oral tradition, and these worked together with the Tier 3 publications to socialize the folk taxonomy.

*4.3.2 Efficacy of Socializing the Folk Taxonomy*

Do these factors form an adequate explanation for the public's resistance to the Copernican taxonomy? The bar of success was set low because there was also a failure of science communication. The textbooks that taught the Copernican taxonomy at the university level did not explain the theory-laden nature of taxonomy, what the corresponding theory was, or why it was important; they simply stated that planets are both primary and secondary with no comment. Even today, there is a lack of understanding of the importance of taxonomy in astronomical science, compared (for example) to biology. This is evidenced by the IAU's vote in 2006 and the poor arguments that were made in its favor at the time.

There was also a lack of new data about the secondary planets to guide development of scientific theory and to create stronger desire for a pragmatically co-evolving taxonomy. Only a few scientists were developing theories of planet formation. Even those who heard the scientific taxonomy may have failed to grasp its importance and may have thought it was disposable.

Scientists were also slipping into language that, while not denying the taxonomy, did not inherently reinforce it. Primary planets were often just called *planets* in contexts where it was not confusing, and secondary planets were more often just called *satellites*. As early as 1767 someone wrote, “When we speak of the Planets without any distinction, we always understand the primary ones” (Anon, 1767). This was not strictly true, since there were many places in the literature where *planet* without distinction meant a secondary planet, but the point is that astronomers were self-aware that they sometimes used a short version “planet” when referring to primary planets. Members of the public may have misunderstood this, thinking that the terminology was indicative of exclusionary classes of objects that aligned with their views carried over from geocentrism.

It seems that even professionals were making concessions to the public’s inadequate view. Scottish writer Ferguson in the 18th century wrote two works aimed at the public, and in both he listed only the primaries as the set of planets (Ferguson, 1753; 1768) with only hints that he thought secondaries are planets, too. However, his work that was aimed at a technical audience clearly used the Galilean/Keplerian taxonomy: “all the Planets both primary and secondary” (Ferguson, 1756).[17] While all the Tier 2 authors prior to 1857 demonstrated that they understood that scientists classified satellites as planets, it seems likely that broad swaths of the public were simply unaware.

The public’s exposure to correct astronomy was limited in another way. Most people during the beginning of the transition period obtained only a primary education and it did not include astronomy at all. Scanning an online repository of primary school textbooks used in North America during that period, we found no books that taught anything about astronomy, not even the names of the planets. The only mentions of planets were in eclectic readers where the various literary authors sometimes mentioned a planet as a poetic symbol of the teleological ideas. As mentioned above, the annual almanacs teaching astrology with their shortened lists of planets were likely the only information most of the public ever received about planets.

There were also complaints about the higher levels of education failing to keep up. In the mid-19th century while knowledge was rapidly expanding, one British pamphleteer asked,

> How much longer are we to continue teaching nothing more than what was taught two or three centuries ago? or, Ought not our highest education to embrace the whole range of our present knowledge? (Zincke, 1850)

Another pamphleteer complained,

> A student may pass through the whole system of education, and eventually leave Oxford with the highest honours and testimonials of merit and proficiency which that University can bestow, and yet…he may be ignorant that…the planets have orbits… (M.E., 1860)

In contrast, the socialization of the folk concept was strong. The almanacs were widespread, and the desire to see orderliness and safety in the cosmos was probably a strong psychological force for its acceptance. Oral tradition likely played an important role. Convincing the public that planets are an orderly system aligned with the assumptions of astrology was probably much easier than what scientists were trying to do, which was to convince the public that planets are messy, disordered, formed by natural process, and numbering in the hundreds (as scientists included the asteroids at that time), moving in multiple types of orbits that are subject to evolution and change. Overall, it seems that the picture we have described here is correct and complete.

[17] Ferguson is an interesting special case that deserves more full treatment to show how his views developed and evolved, straddling science and culture.

## 4.4 What Developed: the Co-Evolving Theory

We are now in a position to assess what exactly was the underlying theory that drove the taxonomical tendencies of the general culture such that moons would end up being considered non-planets. The etymology of the Greek word for planets πλανῆται (wanderers) is often mentioned in modern texts, but the public during the relevant period evidenced no interest in the Greek etymology, and if it had, then moons would be planets because they definitely do wander like the primaries, as the early Copernican scientists repeatedly noted.

The idea that "planets are the bodies that directly orbit the Sun" was the result but it is unlikely to have been the cause: the 19th century public never evidenced concern over what the objects orbit. It was a deep philosophical and theological concern in the 1500s and 1600s while the Copernican Revolution was being fought among intellectuals, but after heliocentrism had been accepted by the public in the late 1700s few people living in the 1800s were worried about implications of the Earth moving, so the inertial center of the Solar System had little contact with their lives. The published record shows that selecting the set of objects that directly orbits the Sun was an outcome of some other consideration, not the driving motive itself.

The interest that is evidenced in the literature was teleological, and in particular it was the idea that planets exist to communicate orderliness in the cosmos: through control over nature as pretended in astrology; and through the mathematical orderliness that reflects general benevolence and design. The literature shows a strong bias toward this in the public, which motivated their development of a simplified, idealized set of planets that conveyed those messages. The Moon had to be ejected from the lists not because it orbits the Earth, but because the public could not maintain the geocentric-like simplicity if they admitted all the other secondaries along with it.

Because orderliness was the driving motive, the classes of celestial objects had to be divided according to the architecture of that orderliness. The top level in the architecture is orbiting the Sun, and the next level is orbiting something that orbits the Sun; a third tier comprises the messy bodies of the system that do not reflect order at all. A taxonomy with those divisions communicates the aspects of planets that most significantly touched peoples' lives. The failure in 19th century astronomy was that the public was not provided deep acquaintance with the geological complexity of planets emerging from planet formation processes, nor were they taught to discount their own perspective bias and thus to recognize that the dynamical orderliness of our Solar System is incomplete, contingent, and temporary. Planets were not incorporated into a web of mutually supporting scientific explanations about nature, so the public chose a non-reductionist theory that had pragmatic value in their lives. They needed a lexical system to discuss that, so the folk taxonomy developed to fill that need. Now it feels natural for the public to see planets as a class defined by orderliness.

# 5. THE GREAT DEPRESSION OF PLANETARY SCIENCE, 1910 – 1955

Bibliometrics show there was exponential growth in the astronomical community from 1700 to the present as shown in Fig. 8. There was a sudden increase in the exponential rates consistent with the 1781 discovery of Uranus by Herschel. Remarkably, beginning ca. 1894 the growth for *astronomy* entered a plateau for 56 years. This may have been extended in part due to the Great Depression and the two World Wars that occurred during period, all of which reduced government spending for science. However, the plateau started before those events, so there must have been other factors. The trend was much worse for planetary science publications than for astronomy, as the bibliometrics for the words *planet* and especially *satellite* entered marked decline beginning ca. 1910. *Satellite* went from a high of 157 publications per year to a low of 62, or a 61% reduction in productivity.

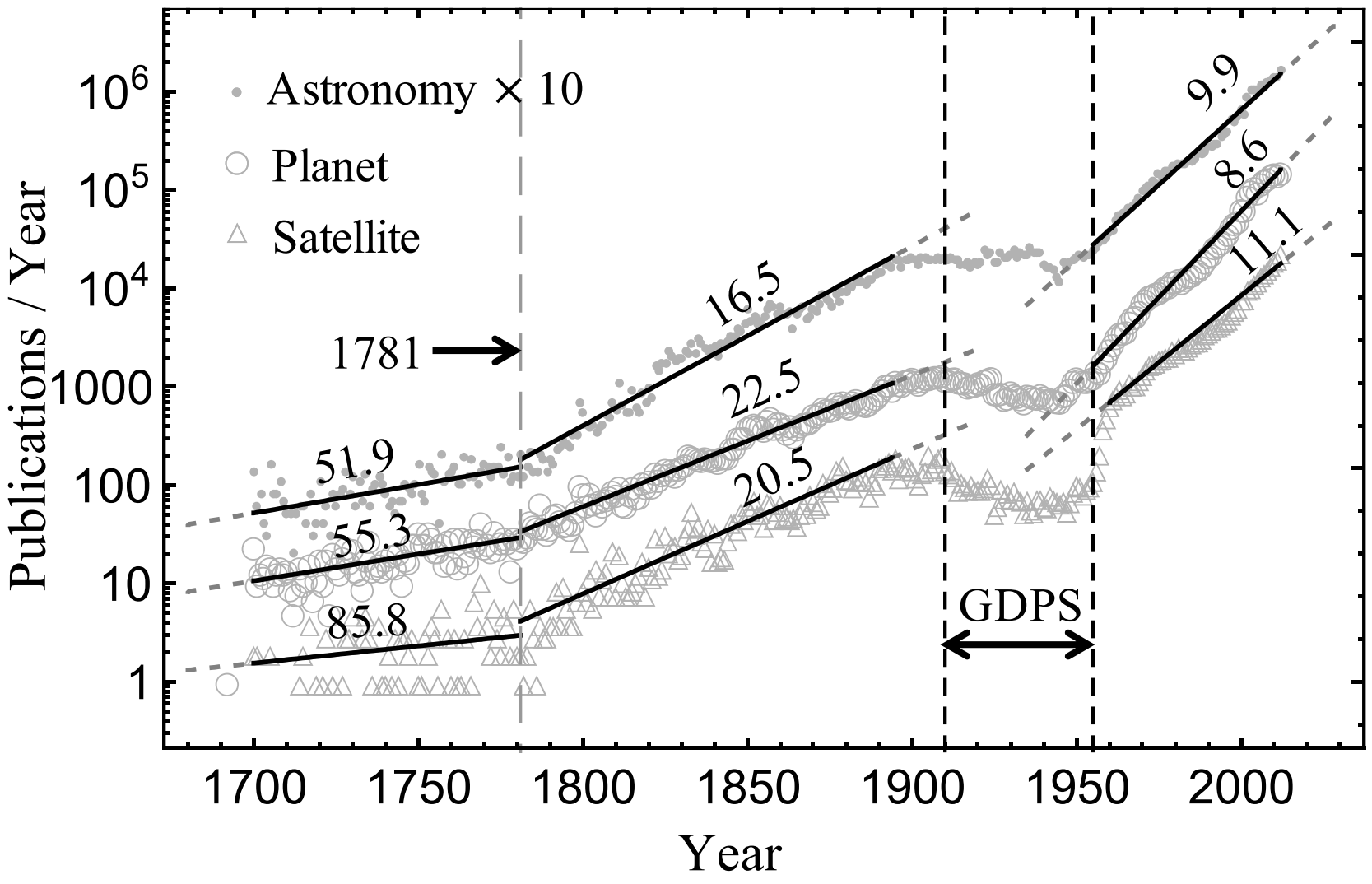

**Figure 8**. Exponential growth in the astronomy and planetary science community indicated by yearly count of publications containing the words *astronomy*, *planet*, and *satellite* as determined by Google Scholar searches restricted to English-only publications. *Astronomy* values were multiplied by 10 for clarity. All three terms change slope close to the 1781 discovery of Uranus, which likely caused the increased interest. Straight lines are best-fitted exponential functions annotated with the doubling period for each in years. The first three were fitted from 1700 through 1781 showing only slow growth. The second three were fitted from 1781 until 1894 showing much faster growth. The three terms generally stop growth ca. 1894, then *planet* and *satellite* began marked decline ca. 1910, the “Great Depression of Planetary Science” (GDPS). The final three exponentials were fitted from 1950 to 2012 except *satellite* was fitted from 1960 to 2012 because it went through an initial growth spurt from 1950 to 1960. The plots end after 2012

since data in Google Scholar appear to be incomplete after that date. (After ca. 1922 it was important to modify the search terms to omit publications that include *gear* or *gears* due to many publications about planetary gears and satellite gears.)

We call this decline the *Great Depression of Planetary Science* (GDPS). The reasons the trend was much worse in planetary science than in astronomy are not specifically known, but we offer the following hypotheses. First, astronomy was already plateaued, so as astronomers put increased focus on other astronomical topics it was at the expense of planets. Second, astronomers were losing interest in planets. It was seen as largely solved, at least to the extent possible with the available instrumentation of the time. Many studies were only calculating ephemerides because the geology of those bodies was not visible. There may have been a perception that planetary studies were out of the mainstream and were attracting eccentric views (e.g., the canals on Mars). The excitement had gone out of discovering new asteroids and satellites, since so many had already been found. Third, there was much excitement in other areas of astronomy about what could now be done with stellar (and eventually galactic) spectroscopy, discovery of other galaxies, relativity, cosmology, stellar evolution, nuclear physics, and radio astronomy. Those going into astronomy as an academic career around then were not generally encouraged to focus on planetary science, and of course an astronomer's training tended not to include the essentials for planetary science like geology, meteorology, and the type of chemistry relevant to planets. It seems likely that during this period astronomers simply lost interest in planets and put their attention elsewhere because planetary studies were stymied by the technology of the times that did not allow the planets and other solar system bodies to be well resolved but astrophysics blossomed owing to numerous new techniques enabled by new detector types in various wavelength ranges.

During this same time, papers that described satellites as a subcategory of planet essentially disappeared. Textbooks stopped teaching that satellites are planets in the 1920s. We have already seen there was a competing folk taxonomy, but we should ask whether there was a legitimate scientific reason for scientists to change their taxonomy, as well. We note for comparison that Metzger et al. (2019) showed that there was a real scientific motive to stop including asteroids as planets in the 1950s, and when scientists published papers that argued those reasons, starting with a pair of papers by the leading planetary scientist Kuiper (1951; 1953), the community abruptly stopped calling them *planets*. This is shown in Fig. 9. Over a period of only 10 years the usage of *planet* for asteroids plummeted from the historic high level of usage that had existed for 150 years to the low level we see today. We investigated whether this is what happened for satellites, too.

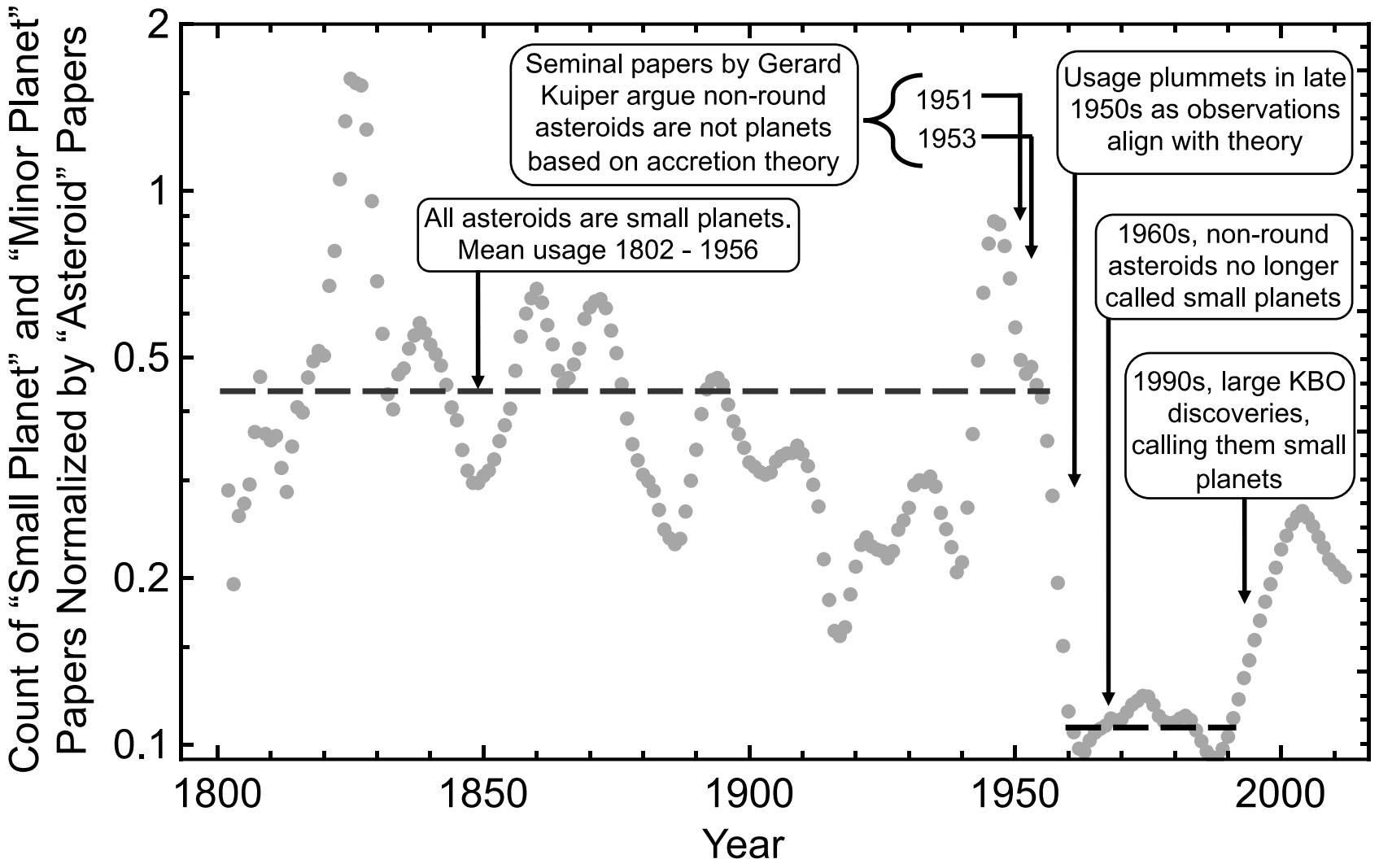

**Figure 9.** Semilog plot of ratio of the number of scholarly papers containing either “small planet” or “minor planet” to the number of scholarly papers containing “asteroid”, with 5-year smoothing for clarity. Horizontal gray lines: guides to the eye for average values over 1801-1956 and over 1960-1991. Adapted from Metzger et al. (2019).

The published literature shows that there was no similar scientific argument for satellites. That is, we found no statements in the literature that satellites ought not to be planets, and we found no arguments that tended to say planets in primary orbits are inherently different types of bodies than those in secondary orbits. In fact, the opposite was the case. Arguments about planet formation compared similarities in the primary and secondary systems for explanatory benefit (Kirkwood, 1860; 1869; Kuiper, 1951). To this day there are no arguments in the literature based on science that moons are an inherently different type of object than primaries. However, there are many arguments in the modern literature for moons to be planets, which we discuss in the next section.

The bibliometric evidence confirms that the change in terminology was not a scientific revolution but a slow process taking centuries to complete. We investigated the intensity of use of the terms *primary planet* and *secondary planet* in the literature. Intensity is defined as the annual count of publications for either term divided by the count of publications using the term *planet* as a proxy for the extensive scale of the planetary literature. Logarithmic curve fitting is needed to match the tails, but in some years these data sets have zero count which is incompatible with logarithms, so it was necessary to convert the data to a cumulative form then fit to that. The fitting form for the differential publication intensity is assumed to be

$$n(y) = n_0 \, e^{-(y-y_0)^\tau / 2\sigma^\tau} \tag{1}$$

with $n(y)$ = the intensity of use, $y$ = the year, and $n_0$, $\tau$, $\sigma$ and $y_0$ = the fitting parameters. Integrating Eq. 1 produces the fitting function for the cumulative publication intensity,

$$\int_{y}^{y_f} n(\zeta)\, \mathrm{d}\zeta = \frac{n_0}{\tau} \left[(y - y_0)E_\beta\left(\frac{(y-y_0)^\tau}{2\sigma^\tau}\right) - (y_f - y_0)E_\beta\left(\frac{(y-y_0)^\tau}{2\sigma^\tau}\right)\right] \quad (2)$$

where $y_f = 2012$ is the final year of the stable Google Scholar data sets, $E_\beta(z) = \int_1^\infty e^{-\beta t}/t^\beta\, \mathrm{d}t$ is the exponential integral function and $\beta = (\alpha - 1)/\alpha$. Fitting for different values of $\tau$, setting $y_0 = 1706$ since it is the start of the data set (the first occurrence of *secondary planet* in English per Google Scholar), and fitting the empirical data with the other parameters finds that $R^2$ is peaked very close to $\tau = 2$ as shown in the inset of Fig. 10. This suggest the empirical intensity of usage was experiencing a Gaussian decline. Setting $\tau = 2$ but allowing $y_0$ and the other parameters to vary finds the best fit for that model ($R^2 = 0.9984$) with $n_0 = 0.0243$, $\sigma = 69.1$ years, and $y_0 = 1714.16$. This fitted curve is shown in Fig. 10 with the cumulative empirical data. These values of $n_0$, $\sigma$ and $y_0$ were then applied to Eq. 1 and compared to the differential empirical data in Fig. 11 but with 9-point smoothing on the empirical data to make it easier to read since the data are noisy. The best fits using $\tau = 1$ and $\tau = 3$ are also shown for comparison, demonstrating that $\tau = 2$ is superior. The differential *primary planet* data are shown in Fig. 12. It is compared with the best fit to the *secondary planet* data using $\tau = 2$ but raised by a factor of 8 (since *primary planet* was used 8 times more often than *secondary planet*). This demonstrates that the usage intensity of both *primary planet* and *secondary planet* were declining at identical rates from about 1800 to about 1955.

This analysis indicates Kepler's primary/secondary planet terminology declined in the English language over a period of centuries in a manner consistent with Gaussian decay, even while it was being taught in the textbooks as the correct terminology. This was a surprising result, and while we do not wish to make too much of it, it may be worth mentioning that a Gaussian decay (versus exponential or some other form) suggests that a purely diffusive process was at work. A conclusion that can be stated more confidently is that there was no forced paradigm shift in the terminology such as occurred for asteroids in Fig 9. The community simply drifted away from Kepler's *terminology* over a long period of time, even while continuing to teach Kepler's and Galileo's *concept*.

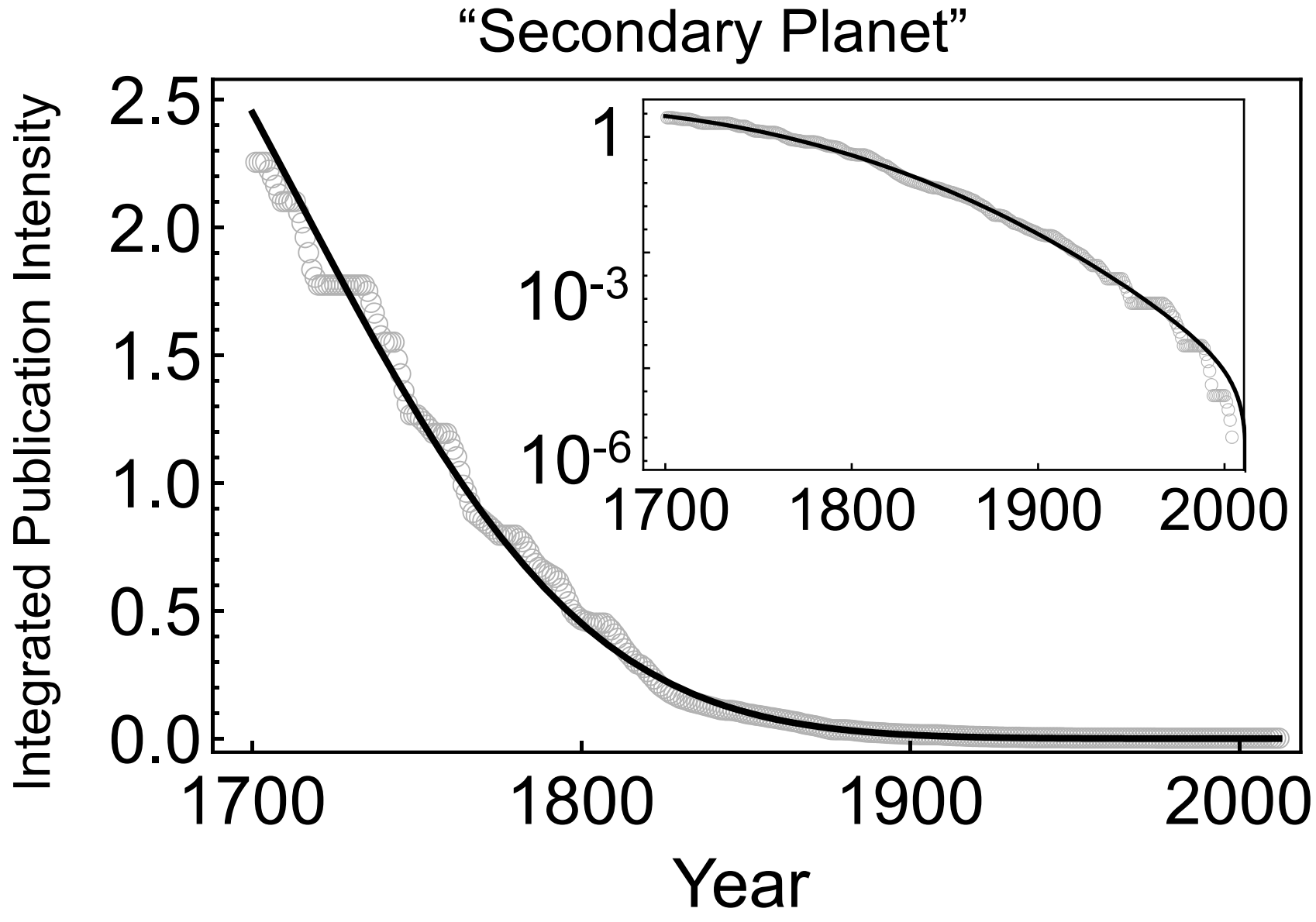

**Figure 10.** Integrated (or cumulative) publication intensity of *secondary planet*. Gray circles: empirical data. Solid line: fitted curve per Eq. 2 setting $\tau = 2$ and finding a least-squares fit between the log of the data and the log of Eq. 2. Inset: the same plot on semilog axes to show the tail.

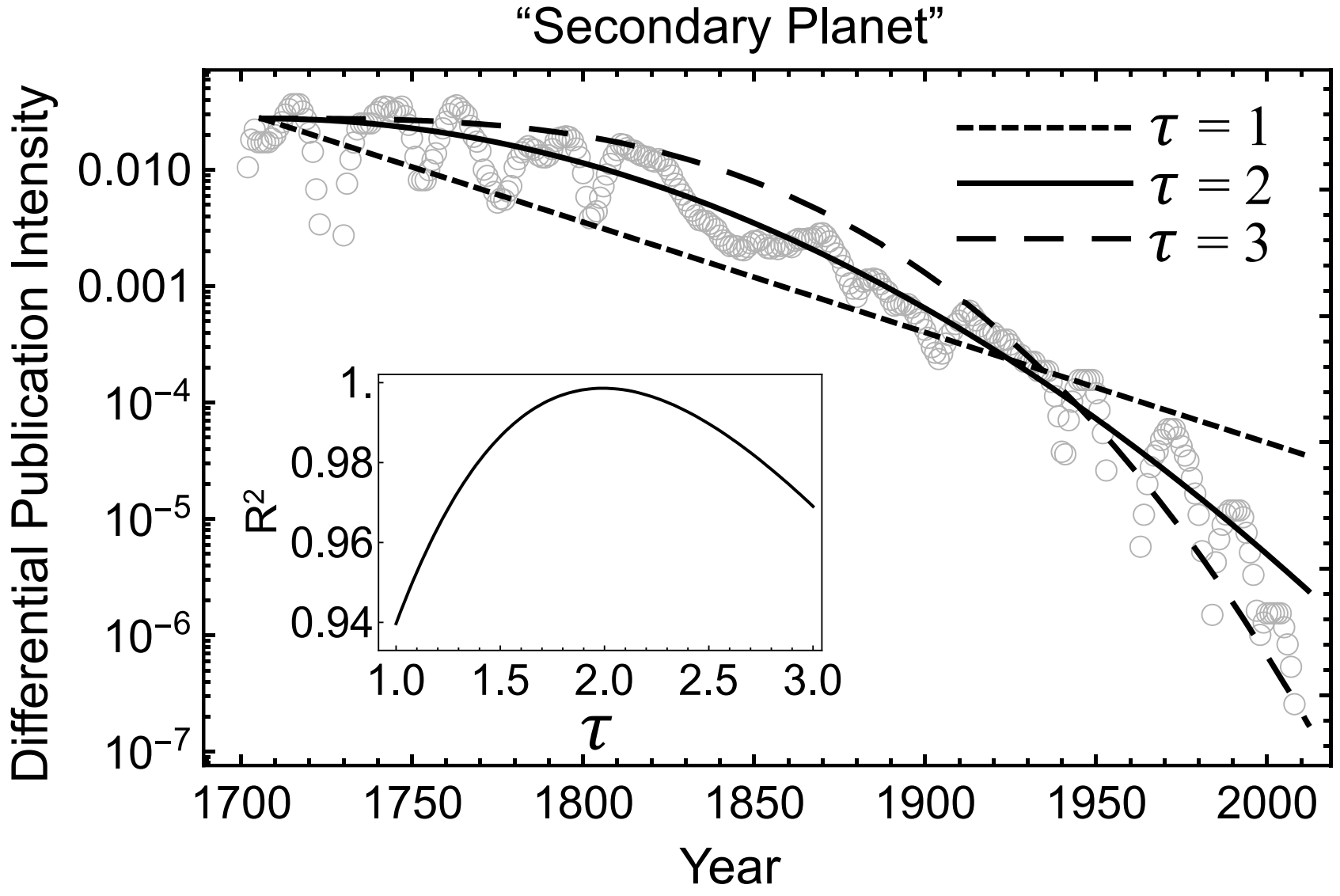

**Figure 11.** Differential publication intensity of *secondary planet*. Gray circles: empirical data with 9-point moving average smoothing to make the plot more readable. Solid and dashed lines: per Eq. 2 using the constants obtained from the fitted curve of the cumulative function of Fig. 10 for cases with $\tau = 1$, 2, and 3. The $\tau = 3$ case might fit better after 1980 when the data began declining faster than Gaussian, but in the relevant period from about 1750 to 1920 the data are clearly consistent with a Gaussian decay, $\tau = 2.$ Inset: $R^2$ calculated from curve fitting with different values of $\tau$ showing it peaks at $\tau = 2$.

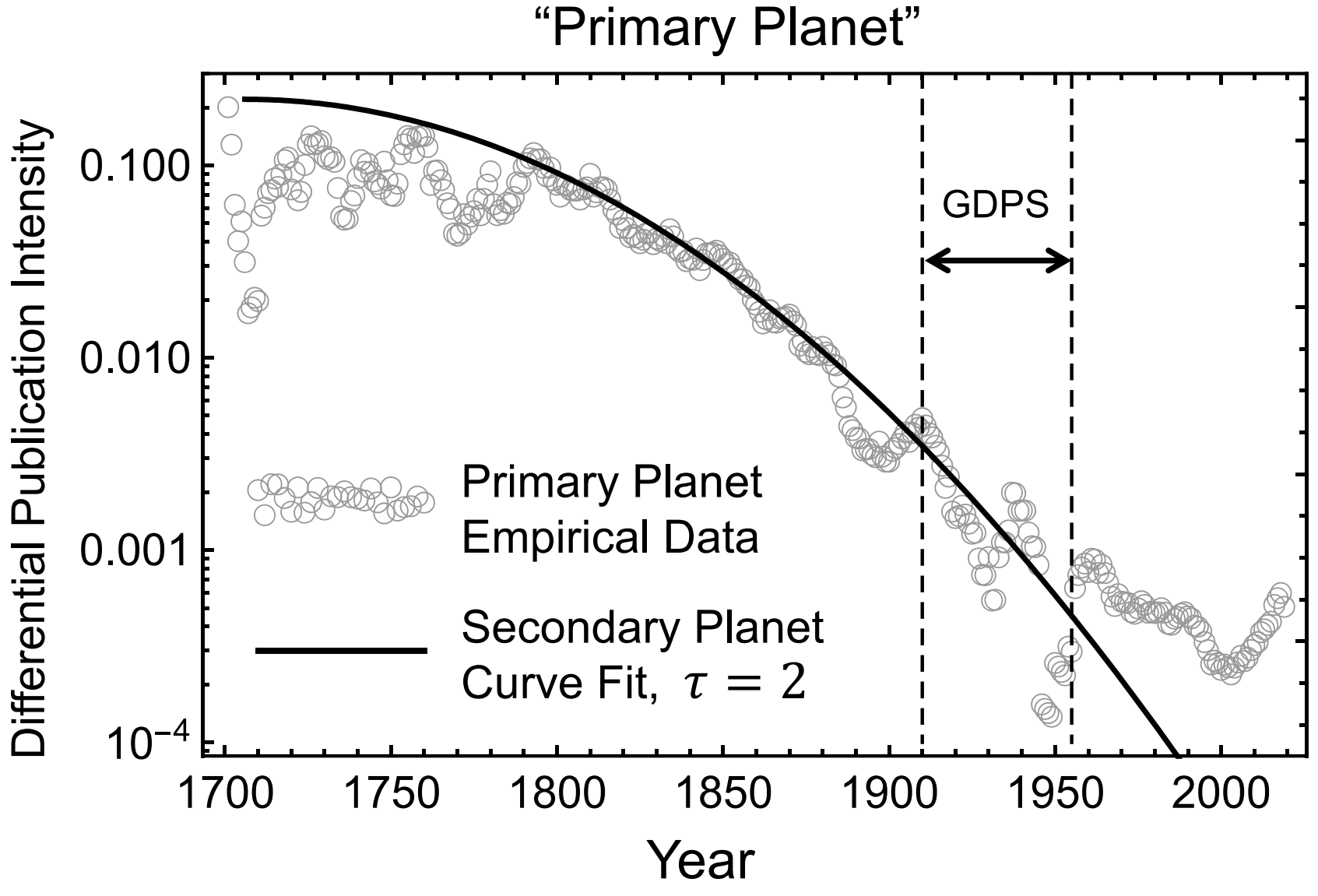

**Figure 12**. Differential publication intensity of *primary planet*. Gray circles: empirical data (no smoothing). The solid curve is the same curve from Fig. 11 with $\tau = 2$ but raised a factor of 8 showing that usage intensity of *primary planet* matches the same declining rate as *secondary planet*, $\sigma \cong 69$ years, over a period of about 250 years. The intensity of *primary planet* usage departs from the Gaussian decline and remains roughly constant after the GDPS.

From this, and from the way satellites were being discussed in the literature, we can infer that scientists were not moving away from the terminology because it was considered wrong or no longer aligned with the best reductionist science, but simply because the term “satellite” was shorter and easier to write than “secondary planet”, and perhaps because “planet” could be understood to mean just the primaries in most cases where there was no chance of confusion. It was apparently a change merely for convenience. Several authors in the literature specifically remarked on the preference for simpler terminology for satellites (Anon., 1767; Leahy, 1910; Chamber, 1911).

The usage intensity of *primary planet* departed from the Gaussian decay trend in the 1950s and thereafter maintained roughly constant usage, which was about when the GDPS ended with the infusion of new geological data from the solar system, as we discuss below. This makes sense, because as scientists began discussing planets with their satellites in more detail, especially planning missions to those systems, they could easily say *satellite* for the latter, but *planet* would be ambiguous in that context so *primary planet* or often just *the primary* (for short) was needed instead. We see examples of this in the literature (e.g., Heward, 1907, p. 227; Holmes, 1914, p. 214; Pickering, 1928), and this is still common usage today. Some publications mentioned only *primaries* and *satellites* (rather than *primary planets* and *satellites*), perhaps leaving ambiguity in the interpretation whether they were discussing primary and secondary *bodies* rather than primary and secondary *planets* (e.g., Antoniadi, 1933; Berlage, 1934; Lyttleton, 1939; Richardson, 1945; Perrine, 1949; Armitage, 1950; Beiser, 1951; Roy 1954).

Since this change in terminology was an unforced drift between synonymous terms, it leaves open the question when and why scientists came to believe along with the folk taxonomy that the current orbital state of a body is an essential part of the *planet* concept. This change could not have occurred earlier than the 1920s because textbooks until then were explicitly teaching that moons are planets. It probably happened before the revival by planetary scientists in calling moons *planet* that was about to occur in the 1960s (see next section) since the new motive to call them planets again would have militated against the change. This puts the timing of the change precisely during the GDPS when astronomers lost interest in planets.

The following factors would also have been at work during that period. First, the change in terminology may have had an influence toward presuming that the dynamical state is an essential part of the *planet* concept. The shorter form of the terminology (planet or primary vs. satellite) that was replacing Kepler's terminology (primary planet vs. secondary planet) was more convenient but no longer served as constant reminder of the taxonomy received from the Copernican Revolution.

Second, before the 1960s scientists had only limited data to inform the geophysical/geological view that satellites are the same type of object as primary planets. Most observational papers before the 1960s were dealing only with ephemerides or global properties that results from unresolved data on a body.

Third, the GDPS itself must have reduced the opportunity for scientists to teach the meaning of the Galilean/Keplerian taxonomy to the next generation, since scientific productivity regarding satellites was diminished by as much as 61% during those 45 years. With 61% less attention on satellites, there would be 61% fewer chances for a professor to emphasize the basis of a taxonomy that unites common geology and geophysics across orbital states, and since mainly just ephemerides were being studied, the chances were already few. The duration of the GDPS was 1.5 professional careers. One or two generations of astronomers failing to learn the reason for the taxonomy would be enough to break the chain of transmission from Galileo.

Fourth and perhaps most importantly, it seems the culture influenced scientists, since the cultural concept is ultimately what astronomers during the GDPS fell into believing. Scientists are of course part of culture and influenced by it (Semin and Gergen, 1990), and the cultural view is what students would learn before going to college. Once you have lost interest in something as a scientific topic, then it is natural to revert to popular culture's usage.

What astronomers did after embracing the folk taxonomy was to fall into the presentism fallacy, doing historical revisionism to support their newly acquired, culturally informed concept of planets based on orbits. We see presentist revisionism happened both with asteroids and with satellites. Considering the asteroids first, we see the record is overwhelmingly clear that scientists considered asteroids to be true planets all the way into the 1950s when there were thousands of known asteroids (Metzger et al., 2019). Asteroids were subclassified by scientists as one of the three types of primary planets. See Kiddle (1870), above, and the following two examples from 1920 and 1941:

> The planets [*Die Planeten*] are decomposed into:
> 1. inner planets [*innere Planeten*], medium-sized, very dense, faintly flattened. This subheading includes Mercury, Venus, Earth and Mars;
> 2. small planets or planetoids [*kleine Planeten oder Planetoiden*], more than half a thousand, all telescopic;[18]
> 3. big planets [*grosse Planeten*], big, not very dense, very flat. Jupiter (the largest of all planets), Saturn (distinguished by its ring), Uranus and Neptune.
> (Kayser, 1920, transl. from German)

> The sun's family of planets is divided according to size into three natural groups: the four major planets, Jupiter, Saturn, Uranus, and Neptune; the five terrestrial planets, Mercury, Venus, Earth, Mars and Pluto; and the countless minor planets or asteroids. The asteroids are so small relative to the other planets that their admission to planetary rank is generally qualified by reservations that are almost apologetic. (Nicholson, 1941)

This last example was only 12 years before the seminal paper by Kuiper (1953) that led to the demotion of the asteroids to non-planet status (see Metzger et al., 2019).[19] The concept of *planet* that included so many small asteroids was quite different than the orderly, cultural concept astronomers hold today. From their modern perspective, modern astronomers can believe easily enough that scientists of the 1800s accepted the first few asteroids as planets, but they find it hard to believe they could have accepted thousands of tiny planets, so they assume that they did *not* accept them. However, when looking at history shortly after the first few were discovered to identify when astronomers stopped considering them planets, modern astronomers cannot find any such event. As a result, they have read into the literature things that are not there, claiming that the decision to start numbering asteroids marks the point where astronomers were not really considering them to be planets anymore. This failure to realize the *planet* concept has changed significantly through history, leading to an interpretation of history based on modern perspectives, is presentism. We see this presentist error about asteroids mistakenly repeated in many recent publications (Hilton, 2001; Soter, 2006; Hughes and Marsden, 2007; Murzi, 2007; Rijsdijk, 2007; Tyson, 2009; Schilling, 2010; Howard, 2012; Bokulich, 2014; Brown, 2017; Masiero, 2017; IAU, 2018; Salyk and Lewis, 2020).

We see the same presentist mistake and historical revision occurred with secondary planets. Modern astronomers now believe that the folk taxonomy has always been the view of astronomers. Since there was in fact no scientific reason for scientists to discard Galileo's *planet* concept and to adopt the folk taxonomy, and since no reason can be found in the historic record, astronomers have to imagine that such a change never occurred. Therefore, they re-wrote history to say that it did *not* occur and that moons stopped being planets with Copernicus. Faced with Galileo and Kepler (and so many others) saying that moons are planets, modern writers have looked upon it as an oddity, an inconsistency in their speech, or a failure to recognize what they had discovered. Dick (2013) wrote, “Terminologically, Galileo was certainly confused.” The presentist error about secondary planets is mistakenly repeated in the following examples:

> If Mercury were orbiting Jupiter rather than the Sun, it would clearly never have been considered a planet (as is the case for Ganymede)... (Basri and Brown, 2006)
>
> 1543…Earth added…sun, moon deleted.…After Nicolaus Copernicus persuaded astronomers that the sun rather than Earth lies at the center, they redefined planets as objects orbiting the sun, thereby putting Earth on the list and deleting the sun and moon. (Soter, 2007)
>
> Thus was born the “solar” system that today we take for granted. But what then becomes of the magnificent seven [i.e., the seven planets of geocentrism]? The Moon and Sun were pulled from their planet status while Earth joined the list…This commonsense reassessment of the word *planet* dropped the number to six… (Tyson, 2009)
>
> …when people thought the moon and the sun revolved around Earth, they were just two of seven "planets"... With the Copernican revolution in the 16th century, the definition of planet changed. They [the planets] were now the things going around the sun. The moon was the one thing going around Earth. (Brown, 2017)
>
> When the Copernican revolution put the Sun at the center of the Solar system, Earth joined the ranks of the planets, and the Sun and Moon were removed. (Masiero, 2017)
>
> In Ancient astronomy, the concept extended to apparently moving or ‘wandering’ celestial bodies, and therefore included both the Sun and the Moon, but not the Earth. With the switch to a Heliocentric system, Planet came to include the Earth, but no longer the Sun, nor the Moon. (Egré and O’Madagan, 2019)

---

[18] Several authors in the 1800s and 1900s called the asteroids "telescopic planets" because they were visible only through a telescope.
[19] Following Kuiper's argument, a larger asteroid like Ceres that was rounded by gravity during formation was a protoplanet and therefore was still a planet. See, for example, Hodgson (1977), Hantzsche (1996), McCord and Sotin (2003; 2005), and McCord (2013).

# 6. RESURGENCE OF SATELLITES AS PLANETS, 1960 – PRESENT

Planetary scientists began rediscovering the usefulness of a taxonomy in which moons are planets at the same time the GDPS was coming to an end. This is also the same period when planetary scientists decided that asteroids are not planets based on new theory and the flood of supporting geological data (Metzger, et al., 2019). Apparently, the ending of the GDPS was brought about in part by planetary spacecraft missions, which were beginning then (e.g., Deutsch, 1960), and the great increase of funding for planetary research that coincided. We see in the literature a sudden surge of papers by planetary scientists referring to large satellites as planets, frequently writing “the planet” instead of the proper name of the satellite. It became common for planetary scientists to call the Earth's Moon a "terrestrial planet" (Head et al., 1977; Warner and Morrison, 1978; Solomon, 1979; Phillips and Lambeck, 1980; Head and Solomon, 1981; Wieczorek, 2007). Moons were compared with primaries when doing comparative planetology, calling the set of objects under consideration "planets" (see examples below and in the supplementary material). This parallels the Copernican Revolution, because scientists looked through better instruments to see geology on other planets just as Galileo looked through a telescope to see mountains on the Moon, and it resulted in a geophysics/geology-based taxonomy that was pragmatic for discussing explanatory theories. Unfortunately, astronomers who are not planetary scientists did not switch back to the geophysical/geological view at this time but continued to embrace the folk taxonomy.

We have done only a limited search since the available automated tools are unable to discern how the word *planet* is used in context; it requires humans reading hundreds of papers to determine and this is very labor intensive. Nevertheless, our limited attempt easily found 149 examples that began in the 1960s. Many more examples can be heard spoken in planetary science conferences. The following examples are generally in papers that describe the complex geology of the satellites. We have added bold as a guide to the eye for the relevant portions of the quotes. We include only a few examples under each heading, but many more are provided in the supplementary material.

1. **Atmospheres**

> Evidence for an atmosphere on **Io**…Thus, if only 10% of the maximum available $CH_4$ froze out during the eclipse over 20% of **the planet**, the resulting layer would be… (Binder and Cruikshank, 1964, bold added)

> ... a way to estimate the meridional heat transport of other **planets**, such as Mars and **Titan**…and potentially to exoplanets too…(Pascale et al., 2013, bold added)

2. **Mantle and Core Physics**

> Subsolidus convection in the mantles of **terrestrial planets**… Each of the **terrestrial planets**, Mercury, Venus, Earth, **Moon** and Mars... (Schubert, 1979, bold added)

> **Europa**: Tidal heating of upwelling thermal plumes and the origin of lenticulae and chaos melting…the viscous response of **the planet** is considered as a first order perturbation of the elastic response to the tidal potential. (Sotin et al., 2002, bold added)

> Ross and Schubert…argue that during the circularization of **Triton**, enough heat was generated to melt **the entire planet**. (Gaeman et al., 2012)

3. **Oceans**

The time of existence of the **Titan's** juvenile ocean was enough for arising of the first protoliving objects. As **the planet** developed through time several energetic processes (irradiation, lightnings, meteoritic and comet impacts) could produce different forms of fixed nitrogen... (Simakov, 2004, bold added)

In Mars, **Europa**, **Callisto**, **Enceladus**, and **Titan**, liquid water may persist to the present […], raising the question of whether **these planets** host active hydrothermal systems similar to those found in Earth's oceans… **Europa** is the most prominent known example of **a small ocean planet**… Considering past heating adds **Ariel**, **Umbriel**, **Dione**, and trans-neptunian objects such as **Charon**, **Ixion**, and **Quaoar** to the list of possible **ocean planets**. (Vance et al., 2007)

4. **Surface Processes**

**Europa** has the largest reflectance amongst the Galilean moons of Jupiter… **this planet**, being comparable with **the Moon** in size… A considerable increase of water fraction in the total mass of **Ganymede** and **Callisto** may take place only if ice VII with a density of 1.66 $Mg/m^3$ exists in the internal central region of **these planets** where pressures reach 50 kbar and more. (Krass, 1984, bold added)

Wrinkle ridge assemblages on the **terrestrial planets**...**Moon**, Mars, and Mercury. (Watters, 1988, bold added)

Equilibrium Conditions of Sediment Suspending Flows on Earth, Mars and **Titan**… to have occurred on other **planets** (e.g., water on Mars and methane-ethane on **Titan**)…critical slopes for equilibrium flow are similar for **planets**. Compared to Earth, equilibrium slopes on Mars should be slightly lower whilst those on **Titan** will be higher or lower for organic and ice particle systems, respectively. Particle size distribution has a similar, order of magnitude effect, on equilibrium slope on each **planet**. (Amy and Dorrell, 2016, bold added)

5. **Biochemistry and Life**

**Titan**: a prebiotic **planet**? (Raulin, 1992)

What makes a **planet** habitable?... All **planets** with substantial atmospheres (e.g., Earth, Venus, Mars, and **Titan**) have ionospheres which expand above the exobase… Europa is probably unique in that it is the only satellite on which a large ocean can be in contact with the silicate layer. On the other moons, the existence of an ocean implies necessarily the occurrence of a very thick high pressure icy layer at the bottom which impedes the contact of the liquid with silicates. But one must keep in mind that, if the amount of water is large (roughly more than 5% wt of **the planet**), the liquid layer is still separated from the silicate by a thick high pressure icy layer... (Lammer, 2009, bold added)

> **Europa**: The search for life on Jupiter's ocean moon…This picture has emerged as we have come to know **Europa as a planet**... Unmasking **Europa** physical processes were actually responsible for the character of **the planet**. (Greenberg, 2010, bold added)

6. **Dynamics Affecting Geophysics**

> Spectrum of the free oscillations of **the Moon**…The model of **the planet** is considered as a sphere of equivalent volume...The amplitudes of these functions are normalized to unity at the surface of the Moon...Since the torsional oscillations are only associated with **the solid planet's** regions...The free oscillations have an important property: they move toward the surface of **the planet** as the oscillation number increases. Therefore, different frequency intervals of the free oscillations are determined by the properties of different regions of **the Moon's** interior. (Gudkova and Raevskii, 2013, bold added)

> **Triton** destroyed and further doomed other moons, and Neptune has in turn doomed Triton. **The planet** is expected to pass by the Roche limit or crash into Neptune's atmosphere in about 3.6 billion years... (Hall, 2016, bold added)

7. **Magnetospheres**

> A magnetic signature at **Io**: Initial report from the Galileo magnetometer…It seems plausible that **Io**, like Earth and Mercury, is **a magnetized solid planet**. (Kivelson, 1996, bold added)

> By synthesizing observations from the Earth, Venus, Mars, and **Titan**, we can qualitatively evaluate how the basic controlling parameters affect the various loss mechanisms through the ionosphere for both **magnetized and unmagnetized planets**…(Lammer, 2009, bold added)

One might criticize these examples as only incidental uses of a single word. However, in each case the authors chose to use *planet* instead of *satellite* knowing it violated the cultural concept, and this indicates the author naturally associates the *planet* concept with these complex geological processes. We also found cases where the author felt it was necessary to apologize for calling these moons *planets*, since they live in an era when the folk taxonomy is believed to be "correct", and yet the geological likeness was so strong and the reductionist value of the association so beneficial they wanted to call them *planets* nevertheless:

> We will generally use the concrete if not always strictly appropriate "planet" rather than "object" or "body" when referring in general to a planet or large moon. (Zahnle et al., 1992)

> Although Europa is formally a moon, planetary scientists lapse into calling a moon a "planet" once its detailed character is revealed to us, and especially if that

> character and the processes involved appear to be planet-like. [This book then proceeds to call Europa a *planet* throughout.] (Greenberg, 2010)

We note that today some authors are using *world* to avoid saying *planet*, thinking they are required to do this to respect the folk taxonomy *planet* concept and the IAU. The word *world* actually has a different, non-geophysical meaning in social sciences, meaning civilization or some center or aspect of human activity. Planetary scientists inherited the application of *world* to planets from the 1600s plurality of worlds, which said all planets were the homes of worlds, which we now know is untrue. Apparently, the use of *world* in planetary science evolved to include not just planets that have intelligent life and civilizations, but any celestial body that would be immersively interesting to humans were we to visit there or to imagine visiting there. See for example Fig. 2 and the referenced book that defines "world" as a place of human experience in contrast to "planet" as an object of geophysical/geological phenomena. As a result, the word *world* is now often applied to bodies as small as asteroid Bennu, far too small to be planets. Replacing *planet* with *world* is also contrary to the planetary science lexicon (see section 7) and to scientific history. It rejects the reductionist taxonomy of the Copernican Revolution and one of its most important insights (see subsection 8.2). It relegates the term *planet* to non-useful, merely cultural status. It communicates bad science to the public (as it would if biologists had abandoned the term "fruit" to the folk meaning). However, the fact these planetary scientists are doing this demonstrates the need for a top-level term that includes both satellites and primaries. Historically *planet* has always been that word. Modern planetary scientists need to be assured that *planet* is indeed the correct term.

# 7. SUPPORTING EVIDENCE

Not just the word *planet*, but the broader lexicon of planetary science is consistent with the idea that large moons are planets and inconsistent with the idea that they are not. The word *planetary* means having the characteristic of, or belonging to, planets. The term *planetary body* developed in the $17^{th}$ and $18^{th}$ centuries as a synonym for *planet* just as *human being* is a synonym for *human*; this was the usage before asteroids were known, when all primaries and secondaries were considered true planets (e.g., Huygens, 1689; Gildon, 1705). After the 1950s when most asteroids were reclassified as non-planets due to their diminutive size and lack of roundness, we continued calling them "planetary bodies" in a reduced sense that they are the leftover building blocks of planets and share some geological processes with planets, so they still provide insight into planets. Satellites both large and small were (and still are) considered *planetary*, and this developed because they, too, were all classified as planets. Thus, the adjective *planetary* has always been essentially geophysical/geological, not defined by a body's orbit, because *planet* has always been essentially geophysical/geological, not defined by a body's orbit, and the terminology only makes sense that way. It would be strange to use *satellitary* instead of *planetary*: a body does not have a "satellitary core", "satellitary ocean", or "satellitary boundary layer" because cores, oceans and boundary layers do not emerge from satellite relationships but from intrinsic physics. If the IAU's definition were consistently applied, the crust of Europa would be a "non-planetary crust" because crusts would not be characteristic of planets and in particular Europa's would be an example of a crust from a non-planet. The same would be true for planetary albedo, planetary body, planetary core, planet formation, planetary Hadley cell, planetary mantle, planetary protection, planetary scientist, and many more.

The “Earth as a Planet” literature also provides a clear window into what planetary scientists consider the hallmark qualities of planethood. A Google Scholar search found many publications discussing “Earth as a planet”, and they were dealing with the topics listed in Table 1. The California Institute of Technology teaches a course called “Earth as a Planet”, the second of four core planetary science courses. Its 2014 syllabus (Caltech, 2014) listed the topics shown in Table 2. These two lists agree that to consider Earth as a planet means to think of it as a complex geophysical/geological system of systems largely isolated from other such bodies by the space between them, or islands of complexity in the cosmos. This is consistent with Galileo’s argument that the Moon and Earth are the same class of object because they share complex geology. Our language reveals that we still think as Galileo did with the same *planet* concept whether we realize it or not.

**Table 1.**
Topics found in the “Earth as a Planet” literature.

| Topics | Topics (cont’d) | Topics (cont’d) |
|---|---|---|
| Geology | Extinction | Marine geology |
| Tides | Mountains | Cyanobacteria |
| Biodiversity | Atmosphere | Deep interior |
| Environmental management | Climate change | Crust |
| Social and societal concerns | Ice | Oceans |
| Origin of life | Vegetation | Ionosphere |
| Age of the Earth | Evolution | Space weather |
| Plate tectonics | Aquatic biology | Comparative planetology |
| Geomagnetism | Soil and rock | Polar wander |
| Radiation balance & albedo | Insects | Geodesy |
| Cloud formation | Environmental physics | Geoengineering |

**Not found: any terms relating to Earth clearing its orbit.**

**Table 2.**
Syllabus topics from Caltech’s “Earth as a Planet” course.

| Topics | Topics (cont’d) | Topics (cont’d) |
|---|---|---|
| Geologic time | The case for plate tectonics | Erosion |
| Stratigraphy | Adiabatic decompression | Rivers |
| Rocks and minerals | Melting reactions | Erosional and depositional landforms |
| Radioactivity and absolute dating | Phase diagrams | Weathering |
| Mineral identification | Volcanism and plutonism | Sedimentation |
| Block diagrams | Ridges | Glacial landforms |
| Seismic waves | Arcs and hotspots | Radiative balance of the earth's atmosphere |
| Earth structure | Sedimentary rocks | Atmospheric structure and circulation; water |
| Isostacy | Rheology and faulting | Earth's radiative budget |
| Igneous rocks | Faults and folds; orogenesis | Ocean structure |
| Geochronology | Metamorphism | Biogeochemical cycles |
| Conduction and convection of heat | Metamorphic rocks | Records of past climate |
| Planetary magnetism | Melting and crystallization | Making geologic maps |
| Seismology | Topography and the orogenic cycle | |
| **Not included: Earth clearing its orbit** | | |

# 8. DISCUSSION

## 8.1 Results of the Literature Review

The literature review found six indicators of the essence of planethood, all of which agree:

(1) Galileo argued the Earth is a planet based on shared intrinsic, geological characteristics with the Moon, which he also deemed a planet. The taxonomy based on intrinsic properties rather than orbits was aligned with key arguments that promoted the Copernican Revolution.

(2) After Galileo, scientists consistently taught that planethood is not derived from what a body orbits or how it orbits. No scientific advancement undermined the *planet* concept of the Copernican Revolution. When, astronomers abandoned it sometime shortly after 1920, what they fell into was merely a folk taxonomy from the 1800s informed by the orderliness of geocentrism, astrology, and theological views of planets that were popular in culture but not used in scientific writings.

(3) When planetary science really began in the 1960s with the new technology of spacecraft missions, the enormous influx of data showed the diversity of complex geology of secondary planets in addition to the Moon, and planetary scientists immediately began calling those large satellites *planet* again in pragmatic usage.

(4) The literature shows planetary scientists have ignored the IAU’s vote in 2006 by continuing to call large primaries, dwarf planets, and large satellites by the common term *planet* into the present.

(5) The lexicon of planetary science implicitly recognizes that complex geology and geophysics, not orbital status, is the hallmark of planethood.

(6) The "Earth as a Planet" literature confirms the understanding of planetary scientists that the essence of planethood is existing as a self-contained geological system of systems, an island of complexity in the cosmos, regardless of orbital parameters.

All these indicators agree that the concept of Galileo and the early Copernicans is still the useful meaning of *planet* in scientific applications, and it is the one that planetary scientists implicitly prefer in published scientific work. In contrast, we searched the literature to find papers that use orbit clearing or present dynamical status as the theoretical basis to hypothesize further properties of those objects (as in the process of Fig. 1), but we were unable to find any such papers.

## 8.2 Geological Complexity As the Essence of a Planet

In principle, geological complexity can be quantified. If, for example, it is defined as the number of novel arrangements of matter departing from randomness, then a primitive meteorite has complexity equal only to the number of distinct minerals it contains. Accreting more meteorites of the same composition does not increase it, so larger asteroids are not automatically more complex than smaller ones. However, when the size is sufficient to effect hydrostatic equilibrium (Sykes, 2007; 2008), or the melting and differentiation of a mantle and core by internal heating beneath the unmolten crust, then the number of unique structures increases (unless the entire body melts and then it is made more uniform, but unique structures increase again with differentiation and recrystallization). If energy and mass transport processes are initiated leading to development of mantle plumes and organized crustal features (for example), they further increase the complexity. A vast diversity of geological structures is known to occur. If an atmosphere and ocean are outgassed and retained, structures in them such as stratification, rotation gyres, jet streams, clouds, and precipitation (to name a few) also occur. Chemical and mineral reprocessing contributes to more diversity. If life emerges then complexity becomes extraordinarily high, related to the number of unique structures in each species and the ecological relationships, and/or to features of the genomes. If civilization emerges, it grows even more. If a planet becomes so large that it is a gas giant, it is unclear how many unique structures survive deep in the body, but unique features of fluid flow and phase changes will still contribute to its overall diversification, including storms and circulation bands, chemical processing, and electromagnetic phenomena. Both terrestrial and giant planets can produce magnetic fields and trap radiation belts. If a planet grows so large that fusion begins, the heat flux has a homogenizing effect, so although a star is still a very complicated phenomenon it would have fewer unique structures than a planet. If the object becomes so large that it collapses into a neutron star or black hole, the diversity of features become far less.

Complexity is therefore maximized in the middle mass range, being much larger in that mass range than in dust specks (extremely simple objects) or in most asteroids (modestly complex objects), but also larger than in a star (modestly complex objects) or in a black hole. Condensates range in mass from small interplanetary dust ~$10^{-16}$ kg to the M87 supermassive black hole ~$10^{40}$ kg. Gravitational rounding occurs at about $10^{20}$ kg and fusion occurs at about $10^{29}$ kg. The log-range of planets defined between rounding and fusion is therefore only 16% of the log-range of condensate masses (see Fig. 13) a small fraction of what nature affords. The universe is estimated to consist of about 68.3% dark energy, 26.8% dark matter, and only 4.9% ordinary matter (NASA, 2013). Of the ordinary matter, 40-50% is the intergalactic medium (Bykov et al., 2008) so only 50-60% is in galaxies. Using the Milky Way as a model for galaxies, its total stellar mass is about $54 \times 10^9\ M_\odot$ (McMillan, 2016), and there are about 200 billion stars with about 10 large planets per star. If we estimate an average planetary mass of 2 $M_\oplus$ based on Borucki et al. (2011) and Zeng et al., (2018), the total planetary mass of the Milky Way is about $12 \times 10^6\ M_\odot$ or about 0.02% of the baryonic mass of the galaxy. Thus, planets comprise about 0.01% of the baryonic mass of the cosmos, 0.002% of the total mass of the cosmos, or 0.0005% of the total energy of the cosmos, yet this tiny fraction we know as planets accounts for the great flourishing of complexity in the cosmos and is key to the existence of life and civilization.

These figures indicate that planets, despite their ubiquity in Milky Way star systems (Petigura et al., 2013; Zeng et al., 2018), are an exceedingly tiny fraction of the material composition of the universe. The fact that such a tiny fraction accounts for the emergence of complexity in the cosmos illustrates the importance of the concept of *planets* in scientific reductionism and philosophy.

It is beyond the present work, but if we were to quantify the number of unique structures in the various bodies as a metric, we expect it would form a curve like the one in Fig. 13. Scientists are already using the abstract concept of quantified planetary complexity to form and test hypotheses (Bartlett, et al. 2020), which demonstrates its importance as a concept.

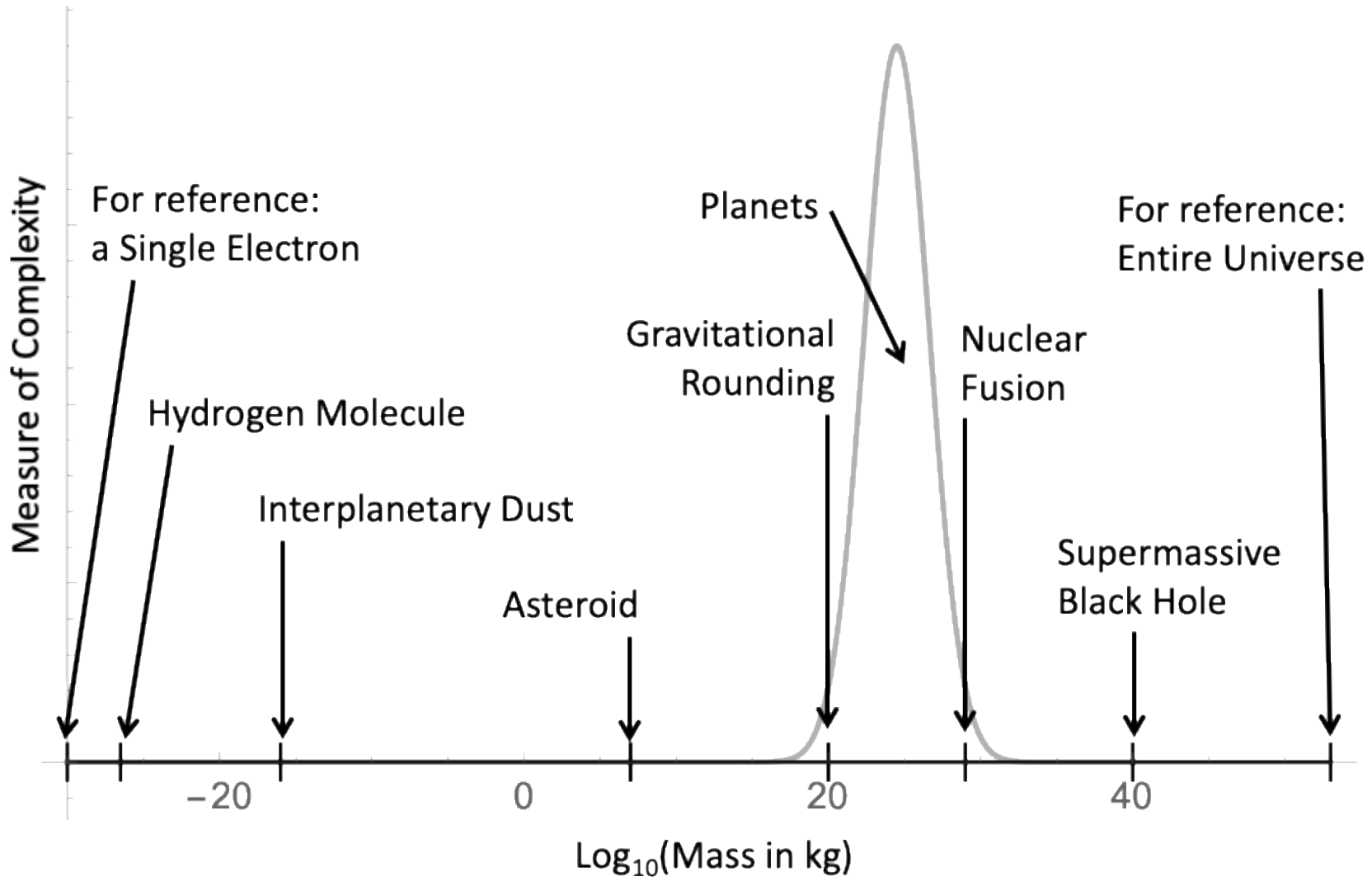

**Figure 13.** (Notional) Complexity of condensates in the cosmos. The measure is expected to be large between gravitational rounding and fusion but small elsewhere.

Perhaps a planet could be defined directly by such a measure of complexity, but for now an appropriate proxy is gravitational rounding at the lower end and fusion at the upper end (Stern, 1991; Sykes, 2007; 2008). These natural transitions of geophysics are meaningful boundaries for all six cases of usage of *planet*, listed above.

## 8.3 Emergence of a Geophysical Planet Definition

Since the 1990s, a formal "Geophysical Planet Definition" (GPD) has been developing to encapsulate this idea. The first step was to define the relationship of planets to small bodies. The discovery of more Kuiper Belt Objects (Stern, 2005) made it important to provide a clearer geophysical statement than had existed since the 1950s when asteroids were reclassified to be non-planets. In 1991, Stern wrote:

> To be considered a planet, an object must
> (a) directly orbit the Sun (or some other star),
> (b) be massive enough that gravity exceeds its material strength (so that the bulk object is in approximate hydrostatic equilibrium), but
> (c) not be so massive that it generates energy through nuclear fusion.
> (Stern, 1991)

See also Stern (2005; 2008 a, b, c), Stern and Levison (2002), and Sykes (2007; 2008). Sykes (2008) stated that this lower mass limit relates to the beginnings of geophysical transport in the body that gives rise to the complex geology. These definitions retained the idea that planets must directly orbit stars. The IAU Planet Definition Working Group incorporated this *planet* concept into its recommendation to the IAU in 2006.

The second step was to delete the requirement that planets must directly orbit a star. This was motivated by the observations of how *planet* is actually used in planetary science (sections 6 and 7). Runyon et al. (2017) wrote,

> A planet is
> (1) a sub-stellar mass body that has never undergone nuclear fusion and
> (2) that has sufficient self-gravitation to be round due to hydrostatic equilibrium
> regardless of its orbital parameters.

Together, these two steps put the dynamical status of a planet back to a secondary tier in the taxonomy the same way that Galileo and Kepler had formulated it. However, the authors proposing the GPD did not know that the GPD is essentially the same as the Galilean/Keplerian *planet* concept, and we did not know that presentism surrounding a folk taxonomy had covered over this fact, until we did the research for this present paper. Therefore, the development of the modern GPD was an unbiased rediscovery of the Galilean/Keplerian taxonomy driven by the modern data.

## 8.4 Significance of the *Planet* Concept

The natural emergence of complexity for a sufficiently large body appears through a series of abrupt transitions, and therefore it is not just an arbitrary boundary like the wavelength we choose to separate reddish purple from purplish red. Things occur in the middle size range of condensates that do not occur anywhere else in the universe, and they include most of the important things in the universe to us: geology, mineralogy, complex chemistry, biology, ecology, history, economics, art, literature, technology, and all the mental activity and dreams of intelligent beings. The fact that this set of transitions exists, and that it can exist only in a specific range of condensates in space, is inherently worthy of discussion and demands a taxonomical category.

This is why the word *planet* is a valuable term in science – one of the most important terms to understand the nature of our existence. This is how it is actually being used by scientists in comparative planetology. This is precisely what Galileo grasped after seeing mountains on the Moon. He was not making the relatively pedestrian claim of the folk taxonomy, that the Earth orbits a luminous body *directly* while the Moon orbits a luminous body only *secondarily*, and thus the Earth and Moon are different in that particular way. He saw instead the truly important thing: that the heavenly bodies have complexity in the form of mountains, and therefore they possess at least some of the processes that make the Earth special as a *world* with life. From this he leapt to the idea that all planets (both primary and secondary) are such bodies. This immediately set off searches for a lunar atmosphere and other elements of geological diversity, and the other Copernicans from the 17th century through the mid- to late-19th century almost universally concluded that many, or perhaps all, the planets are homes to other worlds with life and civilizations (Crowe, 1988). This search for life on planets continues to this day. It has always been complex geology that is the defining concept of a planet.

We hear some make the claim that the *planet* concept is so broad a term because it includes both Mercury and Jupiter that it really lacks any value (Gwynne, 2008). To the contrary, a Venn Diagram (see Fig. 14) showing which planets possess what geological characteristics has overlaps among different subclasses of planets, and shared inclusion percolates across the entire diagram. Every planet is included in multiple subsets, and generalized complexity of geology is the unifying feature. The geophysical boundaries on the upper and lower sizes encompass all these planets and their complexity.

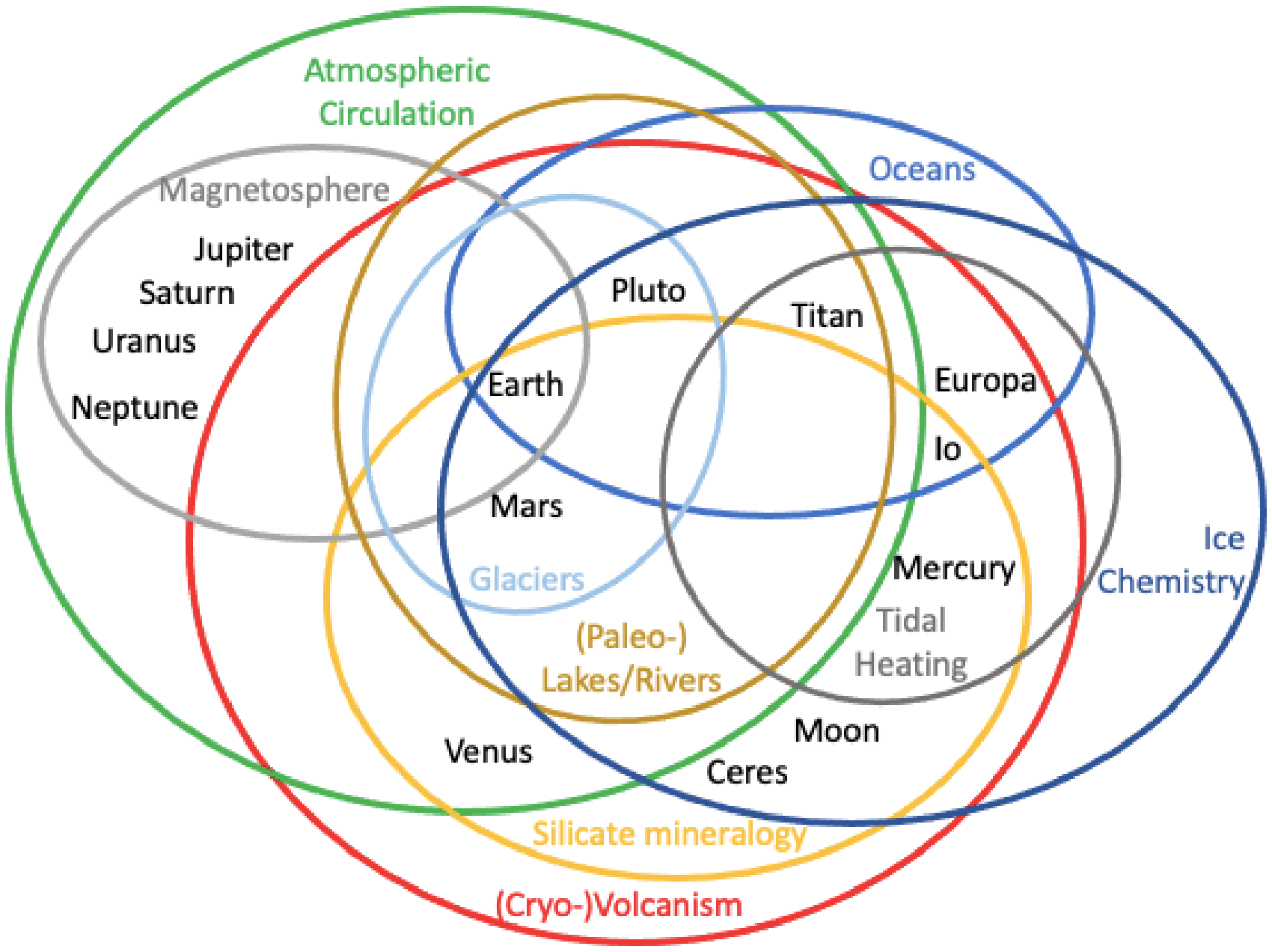

**Figure 14.** (Notional) Venn Diagram of planets with just a few of the various geological characteristics. No planet is contained in all the circles, and some planets like Mercury and Jupiter share few circles with each other, but the overlaps entangle all circles together which argues for the overarching *planet* concept including all these diverse bodies. Planetary diversity therefore does not diminish the concept that geological complexity is their unifying hallmark.

We usually study only a few aspects of geological diversity at a time, but every aspect is important, and their interaction within a planet is central to planetary science. Subcategories like *ice giant*, *terrestrial planet*, *ice dwarf*, *secondary planet*, etc., are adequate to discuss the subgroupings separately, but we need a word for all of them, and historically that word has always been *planet*.

# 9. SUMMARY

Based on these results, the overall evolution of the *planet* concept among both scientists and the general public is summarized in Fig. 15. This is in contrast to the historical revisionist version informed by the presentist fallacy that is commonly taught and believed by many astronomers, which is shown in Fig. 16.

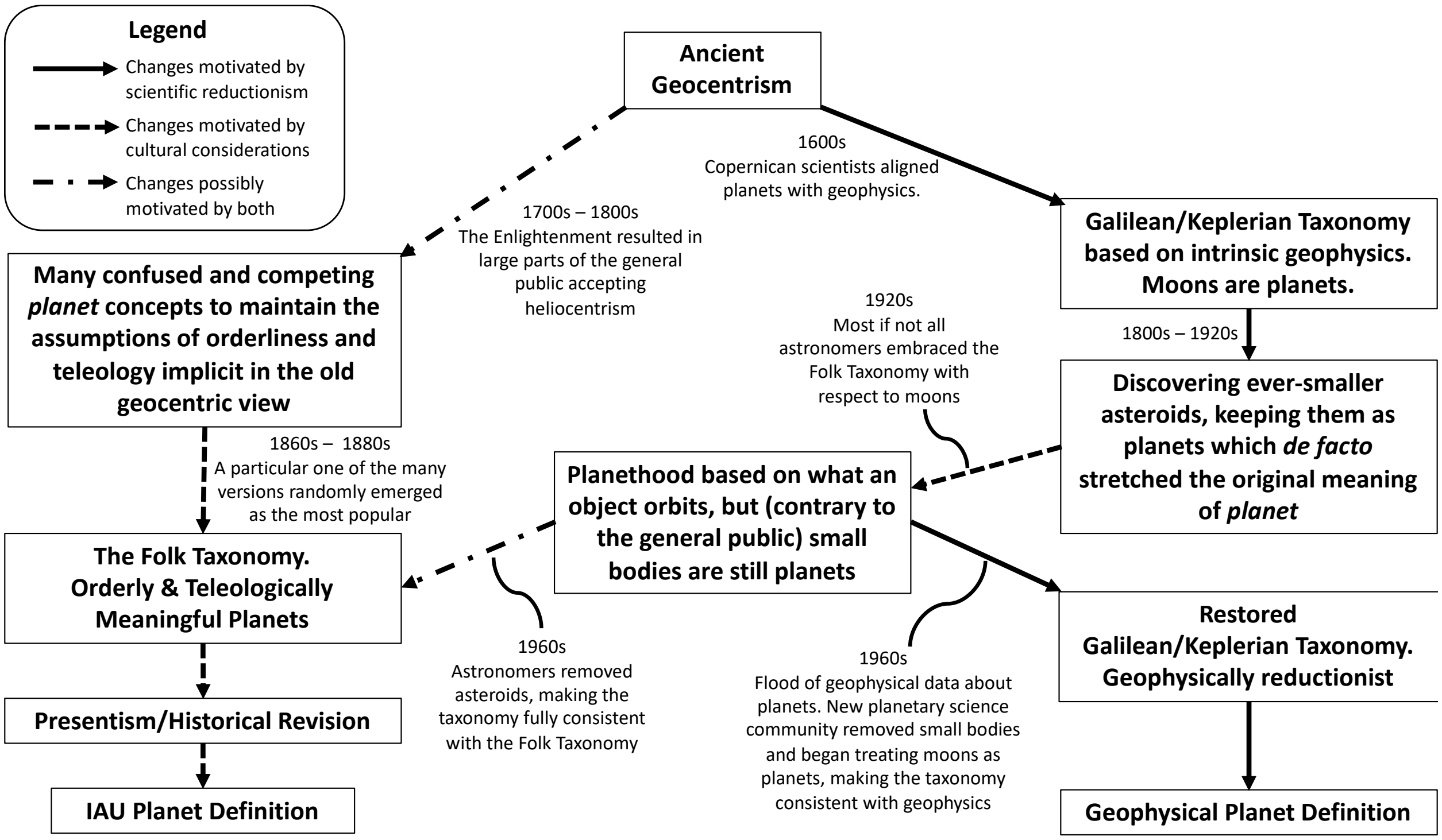

**Figure 15.** The historically accurate evolution of the *planet* concept, including both scientists and culture and how they interacted, as evidenced by the literature.

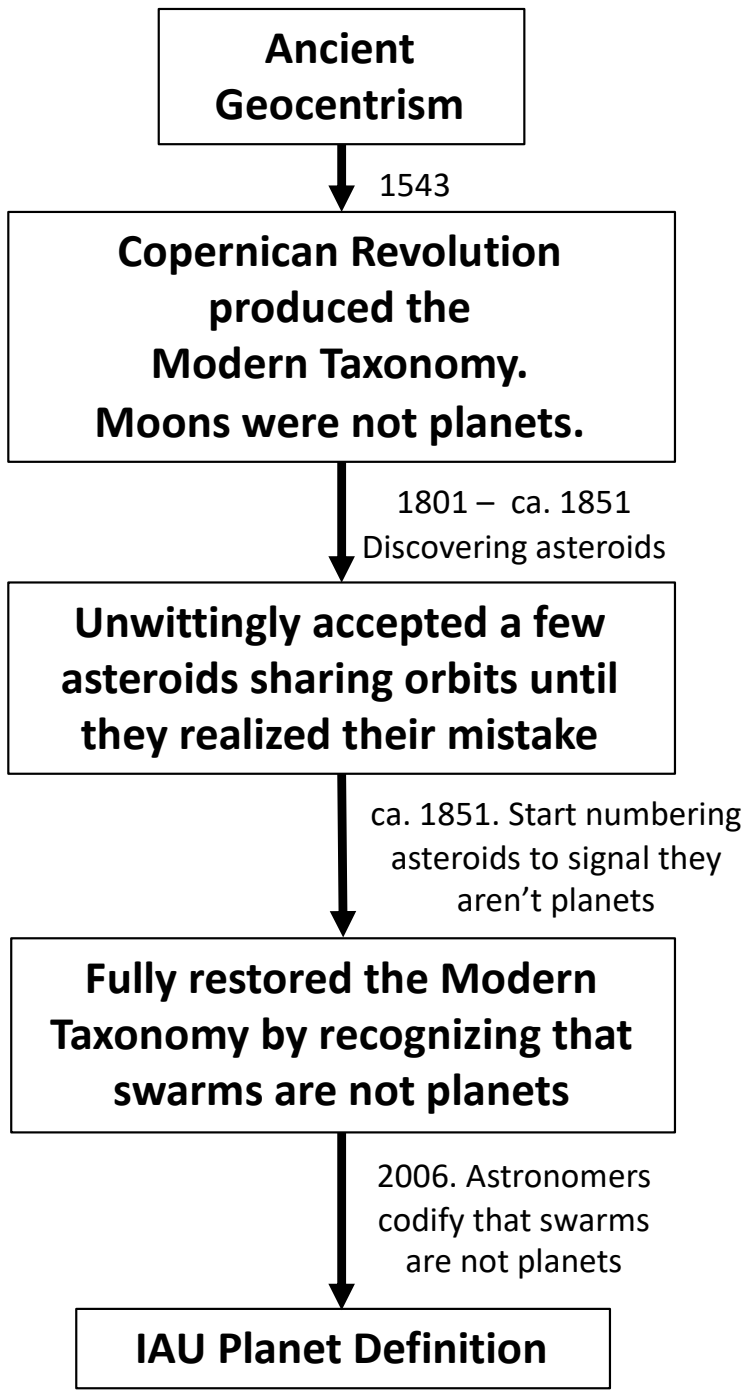

**Figure 16.** The historically revisionist (and demonstrably untrue) version of the evolution of the *planet* concept as commonly taught and believed by many astronomers, contrary to the evidence.

# 10. CONCLUSION

The literature demonstrates that planetary scientists use a concept of *planet* that is fundamentally geophysical/geological, not limited by the current orbital status of a body. A planet is a condensate of intermediate size where physics produces complexity in geology, mineralogy, chemistry, and possibly atmospheres, oceans, magnetospheres, biology, and more, making this *planet* concept not just an arbitrary definition but one of the most important alignments with explanatory insight in all of science.

The Geophysical Planet Definition is a taxonomical formulation that meaningfully encapsulates the concept of geological complexity by saying a planet is any object large enough for gravitational rounding yet not so large as to initiate fusion, regardless of its orbit. This definition and the concept behind it should continue to evolve as science progresses. In formulating this Geophysical Planet Definition, planetary science has come full circle because it turns out that Galileo, Kepler, and the other early Copernicans had the same insight, realizing that planets are *other Earths*, the special places throughout the cosmos where there are complex physical phenomena similar to those of the Earth, up to and including the existence of life and civilization. This led them to define *planets* to include both satellites and primaries, all of which, if large enough, possess this essential character of geological complexity regardless of their orbital location.

For hundreds of years, astronomy textbooks taught this Galilean/Keplerian definition of planets. Unfortunately, sometime after 1920 but before the birth of modern planetary science in the 1960s, the astronomical community had lost interest in planets and during that period abandoned Galileo's key insight, reverting to an astrological folk taxonomy received from culture. That folk taxonomy is the idea that planets are a small set of predictable objects that orbit a common center and depict orderliness and teleological meaning in the cosmos, which the culture carried over from geocentrism and astrology. Astronomers have forgotten that this is the origin of the folk taxonomy and now wrongly teach that this was the concept that scientists developed from the Copernican Revolution.

In recent decades, astronomers began suspecting that Pluto is a member of a belt of smaller objects, and by extrapolating the folk taxonomy they believed this meant that Pluto should not be a "planet". Rather than asking *why* objects should not be planets just because they are in belts (which has no scientific basis or precedent), or where this concept came from, they mistakenly guessed that the concept had historical and scientific precedence in the asteroids (influenced by historical presentism), so they developed mathematical ways to formalize it *a posteriori* via dynamics. This led to an understandable measure of pride among the dynamical community which unfortunately also led to the emotional furor in the 2006 GA when the Planet Definition Committee proposed a geophysical definition for planets without adequate time to sort out the full history.

The concept of orbit clearing or gravitational dominance is indeed important, but it is far less important than the totality of all the other things that should be said about planets, and we found no cases in the literature of scientists using gravitational dominance or orbit clearing to formulate subsequent hypotheses to explain the further nature of these objects (or if such papers exist, they are very few). If gravitational dominance were accepted as the basis of the *planet* concept, it would render the *planet* concept irrelevant for addressing the vast number of other important aspects of planets since it selects a set of bodies that have only gravitational dominance in common leading to no further insight or application. This has led some astronomers to comment that the *planet* concept is not useful in science at all. We have demonstrated from the literature that the opposing geophysical planet concept actually is useful in science and is aligned with both historic and modern usage among scientists. Using the geophysical planet concept with subcategories for the individual features (including gravitational dominance) makes the *planet* concept both useful and deeply insightful for communicating with the public. Because adequate time was not taken to sort these issues, the vote in 2006 created a deeper split in the community.

It is important for every branch of science to reject folk taxonomies and to embrace taxonomies that have the most insightful alignment with explanatory theory, not only to aid scientific practice and to bring consistency to the lexicon, but to communicate deeper scientific insight to the public. Therefore, we should embrace the Geophysical Planet Definition and teach it to students and to the public at large as a correction to the folk idea that what or where an object currently orbits is the definition of what it essentially is. This will require corrections to textbooks and curricula from kindergarten through university. Also, the IAU should rescind their non-scientific definition and stop teaching the revisionist history of its origin. We need to work to bridge the fissure between scientific practice and public understanding and to bring to the public the full benefits of the Copernican Revolution.

# Acknowledgements

We would like to thank Dr. Natacha Fabbri of the Galileo Museum, the Institute and Museum for the History of Science, Florence, Italy, Dr. Carl Brusse, visiting fellow at the Australian National University School of Philosophy, Dr. Victor R. Baker, Regents' Professor of Hydrology and Atmospheric Sciences, Geosciences, and Planetary Sciences at the University of Arizona, and two anonymous reviewers for reading earlier versions of this manuscript and providing helpful comments.